\documentclass[twocolumn, aps, pra, 10pt, showkeys, twocolumn]{revtex4-2}

\usepackage[colorlinks=true, citecolor=blue, urlcolor=blue, linkcolor=magenta]{hyperref}
\usepackage{braket}
\usepackage{amsmath,amssymb,amsthm, mathrsfs}
\usepackage{easybmat}
\usepackage[pdftex]{graphicx}
\usepackage{times,txfonts}
\usepackage{braket}
\usepackage{color}
\usepackage{natbib}
\usepackage{appendix}
\newcommand{\kett}[1]{\left|#1\right\rangle\rangle}
\newcommand{\Tr}{\textrm{Tr}}
\newcommand{\cdott}{\boldsymbol{\cdot}}
\newcommand{\der}{\textrm{d}}
\newcommand{\can}{\textrm{can}}
\newcommand{\cs}{\textrm{CS}}

\newcommand{\W}{\mathcal{W}}
\newcommand{\pow}{\mathcal{P}}
\newcommand{\rtn}{\textrm{RTN}}
\newcommand{\inv}{\textrm{inv}}

\begin{document}
    \title{Finite-Bath Open Quantum Systems: Exact Dynamics}
\author{Devvrat Tiwari\textsuperscript{}}
\email{devvrat.1@iitj.ac.in}
\author{Subhashish Banerjee\textsuperscript{}}
\email{subhashish@iitj.ac.in}
\affiliation{Indian Institute of Technology Jodhpur-342030, India\textsuperscript{}}

\date{\today}

\begin{abstract}
In this work, we introduce a method for deriving exact master equations from the dynamical map for finite open quantum systems coupled to (in)finite reservoirs, using the principle of minimal dissipation. The exact dynamics of the central spin model, which models a finite-bath open quantum system, is developed for two interaction types: Heisenberg and stochastic pure-dephasing interactions. The Heisenberg interaction yields a novel phase-covariant quantum channel in the strong-coupling regime, offering a new platform for studying a range of quantum information protocols. The stochastic pure-dephasing interaction provides the microscopic derivation of the paradigmatic non-Markovian random telegraph noise (RTN) channel, establishing its quantum foundation and offering insight into stochastic couplings. We derive the closed-form master equations for both models. As a demonstration, we explore the thermodynamic performance of these systems as quantum batteries. A direct relationship between quantum heat current and charging power is revealed, and RTN quantum batteries are shown to have advantages in charge storage.
\end{abstract}

\maketitle
\section{Introduction}
A realistic quantum system interacts inevitably with its ambient environment, and its resulting dynamics is addressed by the theory of open quantum systems~\cite{Weiss2011, BreurPoqs, banerjee2018open}. Traditionally, these systems have been analyzed in the weak system-bath coupling regime, characterized by Markovian dynamics~\cite{Louisell1973}. Central to this study is the Gorini-Kossakowski-Sudarshan-Lindblad master equation, a cornerstone for understanding the evolution of these open systems~\cite{GKS_paper, lindblad_1976}. Extending the analysis of open systems to find their dynamics in the strong-coupling regime, which is conducive to memory effects, has been a challenging task~\cite{vega_alonso, Rivas_2014, CHRUSCINSKI20221, Breuer_colloq_nonM, Utagi2020, Tiwari2025, segal_strong_coupling}, as has been the task of obtaining a master equation in this regime~\cite{Hall_andersson_2014}. The structure of the master equation is vital for understanding open quantum systems, as it elucidates processes like pure dephasing and dissipation~\cite{Filippov2020, Haase_2018, Hall_andersson_2014, Smirne_2016}. In quantum thermodynamics, it is indispensable for calculating energy exchange rates and heat currents, while setting the equation to zero reveals the system's steady state~\cite{kosloff_2022, deffner_book, landi_entropy_production, Landi_heat_current_RMP, breuer_effective_Ham1, SB_Thomas_NM_heat_engine, devvrat_thermal_circuit_paper}. The master equation also defines the quantum speed limit, dictating the evolution speed of the system~\cite{Plenio_open_QSL, Davidovich_QSL, Deffner_2017}. In quantum optics, it effectively describes photon emission, absorption, and quantum jumps in atom-cavity systems~\cite{Scully_Zubairy_1997, Agarwal2012, Landi_current_paper_2}. The time-dependent dissipation rates in the master equation allow for the identification of non-Markovian behavior~\cite{Hall_andersson_2014, Utagi2020}. Thus, the master equation serves as a cornerstone for analyzing the dynamical, thermodynamic, and informational properties of open quantum systems across diverse contexts. With this in mind, our work develops a novel technique to derive a unique form of the master equation with minimal dissipation from a quantum system's dynamical map.

The dynamics of an open quantum system strongly coupled to an infinite bosonic bath is complicated to explore owing to its coupling to many environmental modes~\cite{prokofev_stamp_2000}.
An effective route to explore the strongly coupled open quantum systems is through finite-bath systems, where environmental effects are dominated by localized modes~\cite{Finite_bath_esposito, Finite_bath_Strasberg_1}. Recently, these have been in the spotlight in quantum thermodynamics and related domains~\cite{Mahler_2011, Pekola2016, finite_bath_Sanpera, Pekola_finite_bath_2, Tiwari2025}. Dissipative finite baths constructed from bosonic~\cite{Nori_2024} and fermionic~\cite{Nori_2023} pseudomodes, as well as from spin baths~\cite{RTN_dissipative}, have been employed to obtain exact dynamics of open quantum systems interacting with non-Markovian environments. Quantum spin-bath, especially the central spin model, is fundamental to the study of finite-bath open quantum systems~\cite{prokofev_stamp_2000, Arenz_2014, Chiranjib_2017, Devvrat_central_spin1, biswas2025, Lidar_central_spin, Rashid_2025}. These models have become an important topic of research and have been experimentally realized on versatile platforms. Thus, for example, they have found applications in studying spin decoherence using Rydberg atoms, nitrogen-vacancy centers, quantum dots, and in quantum thermodynamics, such as a practical quantum battery using NMR spin systems and thermometry~\cite{Tomza_cs_expt_1, Hanson_cs_expt_4, Cirac_expt_cs_2, central_spin_quantum_dot1, TSMahesh_2022, Campbell_2023}. 

In this work, we develop the exact dynamics of such a system, namely, the central spin model. We investigate two types of interactions: (a). a dissipative evolution generated by a Heisenberg type of interaction, resulting in a new {\it phase covariant} master equation, and (b). a random stochastic coupling between the system and the bath, producing a pure dephasing dynamics generating the random telegraph noise (RTN). This type of evolution has been studied semi-classically using a two-level system subjected to a randomly switching magnetic field~\cite{rtn_model_1, SO_rice, Bergli_2006_2, RTN1_Bergli_2009, Joynt_2008, RTN3_Cai2020}. The noise thus produced is the low-frequency noise and has been implemented experimentally, with a number of applications~\cite{Eli_rtn_2001, Vacchini_rtn_expt, rtn_expt_2024, Paladino_2002, Cialdi_2017}. However, a microscopic origin of the RTN dynamics has been lacking. We show here that the central spin model with a random stochastic coupling can produce such a dynamics. These results are tuned towards applications in quantum thermodynamics, generating battery-charger configurations from the respective interactions. The utility of the master equations obtained in calculating heat current and charging power is demonstrated. 

The structure of the paper is as follows. In Sec. II, we discuss the recipe for obtaining the master equation from the system's dynamics. Section III discusses a general dynamical map for a two-level system, including the dissipative and dephasing dynamics of the central spin model, leading to exact phase-covariant and the RTN model dynamics, respectively. The application of the master equations developed is demonstrated in Sec. IV, followed by the conclusions in Sec. V.

\section{Generic approach: master equation from the dynamics}
Here, we develop a technique to obtain the master equation from the dynamical map of the open quantum system. This is achieved using the principle of minimal dissipation, which ensures that a unique definition of the dissipator of the master equation has the least possible average effect on the physical state, shifting to the highest Hamiltonian (canonical Hamiltonian) contribution~\cite{Hayden_2022, breuer_effective_Ham1} (see~\cite{Colla2025-expt} for an experimental realization and the Supplemental Material (SM)~\cite{supplemental} for a brief theory). In~\cite{Hayden_2022}, it was demonstrated that a minimal norm of the dissipator, for a finite system size, exactly corresponds to taking traceless Lindblad jump operators in the master equation. This can be used to systematically develop and explore the master equation for an open quantum system, as outlined below: 
\begin{enumerate}
    \item Given the map $\Phi_t$ of the reduced state of a system of finite size $d$ (obtained either analytically or numerically), such that $\rho(t) = \Phi_t[\rho(0)]$, the generator $\mathcal{L}_t$ becomes $\dot \Phi_t\Phi^{-1}_t$, which is Hermiticity-preserving and trace-annihilating (HPTA). Note that, the invertibility of $\Phi_t$ is a necessary condition for the time-local generator $\mathcal{L}_t = \dot{\Phi}_t \Phi_t^{-1}$ to be well-defined. More generally, the time-local generator may fail to exist at isolated singular points of $\Phi_t$, a situation well documented in the literature on time-local master equations, see~\cite{CHRUSCINSKI20221} and~\cite{Hall_andersson_2014}. At such points, the generator should be interpreted in a limiting sense, and for numerically obtained maps, pseudoinverse methods can be employed to stabilize the inversion.
    \item Next, the generator $\mathcal{L}_t$ is used to obtain the Choi matrix $\mathcal{C} = \left(\mathcal{L}_t \otimes \mathbb{I}\right)\ket{\psi}\bra{\psi}$, where $\ket{\psi} = \sum_i \ket{ii}$ is the non-normalized maximally entangled state in the space of the generator $\mathcal{L}_t$. 
    \item The Choi matrix is, in general, Hermitian and its spectral decomposition is given by $\mathcal{C} = \sum_j \gamma_j\ket{j}\bra{j}$, where $\gamma_j$ and $\ket{j}$ are its eigenvalues and eigenvectors (normalized), respectively. In step 2, if in the place of the generator $\mathcal{L}_t$, the map $\Phi_t$ is used, we get the usual Kraus operators of evolution.
    \item The eigenvectors of the Choi matrix after de-vectorization (${\rm devec}\left\{\ket{ij}\right\} = \ket{i}\bra{j}$) become the pseudo-Kraus operators $E_k$, and their eigenvalues become the real coefficients $\gamma_k$. Using the $E_K$'s, the generator of the master equation has the decomposition $\mathcal{L}_t\left(\rho\right) = \sum_k\gamma_k E_k\rho E_k^\dagger$. 
    \item Finally, the pseudo-Kraus operators are used to obtain the canonical Hamiltonian $ H^{\can} = -\frac{i}{2d}\sum_k\gamma_k\left[{\rm Tr}\left\{E_k\right\}E_k^\dagger - {\rm Tr}\left\{E_k^\dagger\right\}E_k\right],$ and the minimal dissipator $\mathcal{D}\left(\rho\right) = \sum_k\gamma_k\left[L_k\rho L_k^\dagger - \frac{1}2\left\{L_k^\dagger L_k, \rho\right\}\right]$, where $L_k = E_k - \frac{{\rm Tr}\left\{E_k\right\}}{d}\mathbb{I}$ and thus the master equation 
    \begin{align}
        \frac{d\rho(t)}{dt} = -i[H^{\can}, \rho(t)] + \mathcal{D}\left[\rho(t)\right] = \mathcal{L}_t\left[\rho(t)\right].
        \label{canonical_master_eq}
    \end{align}
\end{enumerate}
See the SM~\cite{supplemental} for an explanation. Notably, the above recipe to obtain the master equation has its roots in the theory of minimal dissipation, developed in~\cite{Hayden_2022}. However, the method for obtaining the pseudo-Kraus operators, the major ingredient in the recipe, from the system's dynamical map is developed in this work, and the constructed master equation is exact for the exact dynamical map. Furthermore, it can be emphasized here that the master equation can be constructed from the numerical solution of the full Hamiltonian dynamics. However, in many realistic scenarios, techniques, such as the hierarchical equation of motion~\cite{tanimura_exact} and reaction-coordinate mapping~\cite{Hughes_reaction_coordinate}, can provide the exact dynamics of the system. A master equation can be constructed using the above formalism in these cases, as well.

Now, we particularize to the master equation for the most general dynamical map for a two-level system.

\section{General dynamical map for a two-level system (TLS)}
We consider the general dynamical map $\Phi_t$ of a two-level system (or a qubit) interacting with its ambient environment (see the SM~\cite{supplemental} for a derivation), with the only constraint that the dynamical map is completely positive and trace preserving (CPTP)
$
    \Phi_t = \sum_{j,k=0}^3 \phi_{j+1\,k+1} \ket{j}\bra{k},
$
where $\ket{0} = \left(1~~0~~0~~0\right)^T, \ket{1} = \left(0~~1~~0~~0\right)^T, \dots$~. The elements of $\Phi_t$ are related by $\phi_{13} = \phi_{12}^*, \phi_{14} = 1 - \phi_{44}, \phi_{31} = \phi_{21}^*, \phi_{32} = \phi_{23}^*, \phi_{33} = \phi_{22}^*, \phi_{34} = \phi_{24}^*, \phi_{41} = 1 - \phi_{11}, \phi_{42} = -\phi_{12}$, and $\phi_{43} = -\phi_{12}^*$. 
The corresponding matrix form of the Lindblad superoperator using $\mathcal{L}_t = \dot\Phi_t\Phi_t^{-1}$ is given by 
$
    \mathbb{L}_t = \sum_{j,k=0}^3 l_{j+1\,k+1} \ket{j}\bra{k},
$
where $l_{13} = l_{12}^*, l_{31} = l_{21}^*, l_{32} = l_{23}^*, l_{33} = l_{22}^*, l_{34} = l_{24}^*$, and $l_{4\,k+1} = -l_{1\,k+1}$. 
The relationship between the elements of the matrix $\mathbb{L}_t$, the matrix form of $\mathcal{L}_t$, and $\Phi_t$ is provided in the SM~\cite{supplemental}. It can be observed in the above equation that the last row is the negative of the first row. This affirms the null-space of a matrix and guarantees the existence of a steady state of the system. Now, the Choi matrix can be obtained using step 2 outlined above and is given by
$
    \mathcal{C} = \begin{pmatrix}
        A&B\\
        B^\dagger &-A
    \end{pmatrix},
$
where $A = \begin{pmatrix}
    l_{11}&l_{12}\\l_{12}^* &l_{14}
\end{pmatrix}$, and $B = \begin{pmatrix}
    l_{21}&l_{22}\\l_{23} &l_{24}
\end{pmatrix}$. The eigenvalues and eigenvectors of this matrix determine the dissipation coefficients $\gamma_k$ and pseudo-Kraus operators $E_k$, respectively. From this, the master equation can be obtained.  

Though this method is applicable for any two-level system dynamics, here we particularize this technique to obtain the exact master equation for the finite-bath central spin model for two different interactions: a Heisenberg XX interaction with constant interaction strength and a stochastic time-dependent interaction, respectively. The former provides a new {\it Phase Covariant} master equation, while the latter elucidates the microscopic origins of the well-known RTN model.

\subsection{Dissipative dynamics of the central spin model: an exact phase covariant evolution}
In the central spin model, a spin-$1/2$ particle is surrounded by $N$ other spins, and the central spin interacts uniformly with the bath spins~\cite{Devvrat_central_spin1, Tiwari2025}. Here, the total Hilbert space of all $N$ bath spins is conveniently reduced to an $(N+1)$-dimensional space using the collective angular momentum operators $J_k = \frac{1}{2}\sum_j\sigma^k_j$ (for $k = x, y, z$), where $\sigma^k$'s are the Pauli matrices. The Hamiltonian ($\hbar=1$) of the total system is given by
\begin{align}\label{eq_CS_Ham}
    H &= H_S + H_B + V=\frac{ \omega_0}{2}\sigma^z + \frac{ \omega}{N} J_z + \lambda (\sigma^xJ_x+\sigma^y J_y),
\end{align}
where $\lambda = \epsilon/\sqrt{N}$. $\omega_0$ is the transition frequency of the central spin-$1/2$ particle, $\epsilon$ is the interaction strength between the system and the bath, with $\sqrt{N}$ being the scaling factor, and $\omega/N$ is the scaled frequency of the bath. The evolution of the composite system for arbitrary initial states of the system $\rho_S(0)$ and bath $\rho_B(0)$ by means of the global unitary $U = e^{-iHt}$ is given by 
$
    \rho_{SB}(t) = e^{-iHt}\left\{\rho_S(0) \otimes\rho_B(0)\right\}e^{iHt}.
$
The reduced state of the central spin system is given by
$
    \rho_S(t)={\rm Tr}_B\left[ e^{-iHt}\left\{\rho_S(0) \otimes \rho_B(0) \right\}e^{iHt}\right],
$
where $H$ is the total Hamiltonian. Here, we consider the Gibbs state as the initial state of the bath, which is given by
$
    \rho_B(0)= \frac{e^{-\beta H_B}}{Z}  =\frac{1}{Z}\sum_{n=0}^{N}e^{-\frac{\beta\omega}{2}\left(1-\frac{2n}{N}\right)}\ket{n}\bra{n},
$
where $\beta$ is the inverse temperature, $\ket{n}$ is the standard computational basis, and $Z = \sum_{n=0}^{N}e^{-\frac{\beta\omega}{2}\left(1-\frac{2n}{N}\right)}$. 
Using spectral decomposition of the total Hamiltonian, we find the exact dynamics of the central spin, see the SM~\cite{supplemental}. Considering an arbitrary initial state of the system $\rho_S(0) = \begin{pmatrix}
    \rho_{00}(0) && \rho_{01}(0)\\\rho_{10}(0) && \rho_{11}(0)
\end{pmatrix}$, the density matrix dictating the evolution of the system is given by
$
    \rho_S(t)=\begin{pmatrix} \rho_{00}(t)&&\rho_{01}(t)\\
        \rho_{01}^*(t) &&  1 - \rho_{00}(t)
    \end{pmatrix},
$
where $\rho_{00}(t) = \alpha_t\rho_{00}(0)+\eta_t\rho_{11}(0)$, and $\rho_{01}(t) = \delta_t\rho_{01}(0)$.
The forms of the elements $\alpha_t, \delta_t$, and $\eta_t$ are provided in the SM~\cite{supplemental}. The elements of the dynamical map $\Phi_t$ for this model are $\phi_{11} = \alpha_t, \phi_{44} = 1 - \eta_t$, and $\phi_{22} = \delta_t$; all the other $\phi_{jk}$'s are zero. This places the present calculation in a broader perspective of a general TLS.
Consequently, the elements of $\mathbb{L}_t$ become $l_{11} = \zeta_t = \left[\dot\eta_t(\alpha_t-1) - \dot\alpha_t(\eta_t - 1)\right]/(\alpha_t - \eta_t)$, $l_{14} = \Gamma_t = \left[\alpha_t\dot\eta_t - \eta_t\dot\alpha_t\right]/(\alpha_t-\eta_t)$, and $l_{22} = \Theta_t = \dot \delta_t/\delta_t$; all other $l_{jk}$'s are zero. Notably, the fixed point in the dynamics of the central spin model is given by the eigenvector corresponding to the zero eigenvalue of the matrix $\mathbb{L}_t$. The analytical form for this fixed point is given by 
$\rho_S^{\rm SS} \propto \begin{pmatrix}
    -\Gamma_t/\zeta_t & 0\\0&1
\end{pmatrix}$.
Furthermore, the elements $l_{11}, l_{22}$, and $l_{14}$ constitute the Choi matrix $\mathcal{C}$ for this model. The eigenvalues of this Choi matrix are $\lambda^{\cs}_1 = \Gamma_t$, $\lambda^\cs_2 = -\zeta_t$, $\lambda^\cs_3 = \left[\zeta_t - \Gamma_t - \sqrt{(\Gamma_t + \zeta_t)^2 + 4|\Theta|^2}\right]/2$, and 
$\lambda^\cs_4 = \left[\zeta_t - \Gamma_t + \sqrt{(\Gamma_t + \zeta_t)^2 + 4|\Theta|^2}\right]/2$. The corresponding pseudo-Kraus operators obtained from the eigenvectors of the Choi matrix are $E_1^\cs = \begin{pmatrix}
    0&1\\0&0
\end{pmatrix} = \sigma^+$, $E_2^\cs = \begin{pmatrix}
    0&0\\1&0
\end{pmatrix} = \sigma^-$, $E_3^\cs = \frac{1}{\sqrt{1 + |\xi_3(t)|^2}}\begin{pmatrix}
    \xi_3(t)&0\\0&1
\end{pmatrix}$, and $E_4^\cs = \frac{1}{\sqrt{1 + |\xi_4(t)|^2}}\begin{pmatrix}
    \xi_4(t)&0\\0&1
\end{pmatrix}$, where $\xi_{3, 4}(t) = \left[\lambda^\cs_{3, 4} + \Gamma_t\right]/2\Theta^*_t$. These operators form the canonical Hamiltonian and the minimal dissipator for the central spin model. The canonical Hamiltonian thus formed is given by $H^\can_\cs(t) = \frac{-\Im(\Theta_t)}{2}\sigma^z$, where $\Im(z)$ denotes imaginary part of $z$. Further, the Lindblad operators $L_k$'s, using the pseudo-Kraus operators, constituting the minimal dissipator are given by $L_1 = \sigma^+$, $L_2 = \sigma^-$, and $L_{3, 4} = \frac{\xi_{3, 4}-1}{2\sqrt{1 + |\xi_{3, 4}(t)|^2}}\sigma^z$. Substituting the canonical Hamiltonian and the minimal dissipator in Eq.~\eqref{canonical_master_eq} and after algebraic simplifications, the exact master equation for the evolution of the central spin system interacting with the spin bath becomes
\begin{align}\label{eq_CS_master_eq}
    \frac{d\rho_S(t)}{dt} =& -i\left[H^\can_\cs(t), \rho_S(t)\right] \nonumber \\
    &+ \sum_{k = +, -, z}
    \nu_k\left[\sigma^k\rho_S(t)\sigma^{k,\dagger} - \frac{1}{2}\left\{\sigma^{k,\dagger}\sigma^k, \rho_S(t)\right\}\right],
\end{align}
where the rates $\nu_+ = \Gamma_t, \nu_- = -\zeta_t$, and $\nu_z = \frac{\zeta_t - \Gamma_t - 2\Re(\Theta_t)}{4}$ correspond to energy gain, loss, and dephasing, respectively.
The above master equation is phase covariant, governing an evolution that follows $e^{-i\sigma^z\varphi}\Phi_t[\rho_S(t)]e^{i\sigma^z\varphi} = \Phi_t\left[e^{-i\sigma^z\varphi}\rho_S(t)e^{i\sigma^z\varphi}\right]$. Phase covariance implies a specific rotational symmetry of the system’s dynamical map, ensuring that the evolution of populations (energy exchange) and coherences (dephasing) is dynamically decoupled. Consequently, the environment causes the decay of coherences but cannot spontaneously generate them from diagonal states~\cite{Filippov2020}.
This highlights the advantage of writing down the master equation from a map, as it clearly illustrates the physical processes occurring within a system. This is a major result of this work, as obtaining it from the exact dynamics of the central spin model establishes that this symmetry is an intrinsic feature of the many-body interaction itself. The phase covariant dynamics of the model is also guaranteed because of the following symmetries. The conservation of total $z$-angular momentum, $[F_z, H] = 0$ with $F_z = \sigma^z/2 + J_z$, combined with the fact that the thermal bath state commutes with $J_z$, leads to the dynamical map commuting with $z$-rotations on the system. At the level of the dynamical map, this symmetry is therefore a structural feature emanating from the Hamiltonian and the initial bath state, and the phase covariant form of the master equation follows from it. The novelty of our result lies in two aspects: (a). \textit{Exact, non-perturbative rates:} The explicit closed-form expressions for the time-dependent rates $\nu_+(t)$, $\nu_-(t), \nu_z(t)$ constitute the physically meaningful content of the phase covariant master equation. These rates, derived from the exact finite-bath dynamics at arbitrary temperature and coupling strength, are new and non-trivial. In contrast, previous phase covariant master equations~\cite{Haase_2018, Smirne_2016} were obtained in the weak-coupling regime after applying the secular approximation, and~\cite{Vacchini_2010} used a fourth-order perturbative expansion, of the spin-boson model. Our rates are exact for any coupling strength. (b). \textit{Finite-bath and strong-coupling validity:} The demonstration that phase covariance persists exactly in the strong-coupling, finite-bath regime, where perturbative treatments break down, and memory effects are significant, is itself noteworthy. The comparison with the second-order perturbative canonical Hamiltonian shown in the SM~\cite{supplemental} illustrates that perturbative approaches can significantly misrepresent the dynamics in this regime.

Further, the exact evolution obtained here can be written in the operator-sum representation, $\rho_S(t) = \sum_{j}K_j(t)\rho_S(0)K_j(t)^\dagger$, where $K_j$'s are the Kraus operators~\cite{KRAUS1971} satisfying $\sum_{j}K_j(t)^\dagger K_j(t) = \mathbb{I}_S$. The Kraus operators derived for this evolution are presented in the SM~\cite{supplemental}. These Kraus operators characterize a novel quantum channel and open the gates for a number of applications in quantum metrology, channel capacity, information, and quantum thermodynamics, among others.

It is pertinent to mention here that attempts have been made to derive the master equation for a central spin model, modulo restrictions: in~\cite{Chiranjib_2017}, the Holstein-Primakoff transformation~\cite{Holstein-Primakoff_paper} was used, which is exact only for an infinite bath, while in~\cite{Samyadeb_2017}, an infinite temperature reservoir was assumed. The master equation derived here is exact for a finite bath and for a reservoir at any temperature. 
The advantage of the approach developed in this work lies in its versatility, making it applicable to more intricate situations, like scenarios involving multiple spins within a system or complex system Hamiltonians that incorporate tunneling terms. 

Further, perturbation techniques have been used in recent times to study the dynamics, canonical Hamiltonian, and minimal dissipator of an open quantum system~\cite{colla2025_perturbation, Hayden_2022}. Applying the perturbation technique up to second order for the central spin model, shown in the SM~\cite{supplemental}, the canonical Hamiltonian is given by $\tilde H^\can(t) = H_S +\lambda^2 H^{(2)}_S(t)$, where $H^{(2)}_S(t) = c(t)S_1\left[e^{\beta\omega/N}\ket{0}\bra{0} - \ket{1}\bra{1}\right]$ with $c(t) = N\left[\cos\left(\omega t/N\right) - \cos\left(\omega_0 t\right)\right]/(N\omega_0 - \omega)$ and $S_1 = -e^{\beta\omega/(2N)}\frac{\left[(N+2)\sinh(\beta\omega/2) - N\sinh\left(\beta\omega(N+2)/2N\right)\right]}{\sinh(\beta\omega(N+1)/2N)(e^{\beta\omega/N} - 1)^2}$. As demonstrated in the SM~\cite{supplemental}, the canonical Hamiltonian resulting from the second-order perturbation significantly diverges from the exact canonical Hamiltonian derived here, illustrating that the perturbative approach might not fully account for the accurate dynamics of the central spin model in the strong coupling regime.

\subsection{Dephasing dynamics of the central spin model: RTN model} Here, we consider a variation in the central spin model to model pure dephasing dynamics. The system and the bath Hamiltonians remain the same as in Eq.~\eqref{eq_CS_Ham}. The interaction Hamiltonian is taken to be $H_{SB} = \frac{\epsilon(t)}{\sqrt{N}}\sigma^zJ_z$, where $\epsilon(t)$ is a time-dependent interaction parameter, which originates from a stochastic process. It can be readily verified, as outlined in the SM~\cite{supplemental}, that for the combined system-bath initial state $\rho_{SB}(0) = \rho_S(0)\otimes\rho_B(0)$, the reduced dynamics of the system is given by 
$
    \rho^{01}_S(t) = e^{-i\omega_0t}\Lambda(t)\rho^{01}_S(0)
$, where the dephasing factor is $\Lambda(t) = \Tr\left[e^{-\frac{2i}{\sqrt{N}}J_z\int_0^t \epsilon(s)ds}\rho_B(0)\right]$ and the population terms of the density matrix remain unaffected. For an initial ground state of the bath, the factor $\Lambda(t)$ becomes $e^{i\sqrt{N}\int_0^t\epsilon(s) ds}$. Further, $\epsilon(t)$ being stochastic, the dephasing factor $\Lambda(t)$ can be obtained as an average over the corresponding stochastic distribution $\left\langle e^{i\sqrt{N}\int_0^t\epsilon(s) ds}\right\rangle_\epsilon$. We assume $\epsilon(t)$ switches between $\pm b = \pm\epsilon\sqrt{N}$ randomly with rates $\gamma_{\pm}$ and $\gamma_{\mp}$, where $\gamma_{\pm}(\gamma_{\mp})$ denotes a switch from $+b(-b)$ to $-b(+b)$. This is the well-known \textit{random telegraph process} (RTP)~\cite{SO_rice, RTN1_Bergli_2009, RTN3_Cai2020}, resulting in the RTN. This process has been extensively studied in the context of decoherence produced by a discrete environment on a charge Josephson qubit~\cite{Paladino_2002}, and is directly observed in single-electron tunneling devices~\cite{Zorin_1996, Nakamura_2002}. A symmetric RTP has rates $\gamma_{\pm} = \gamma_{\mp} = \gamma$. Traditionally, an RTP process is modelled by a qubit experiencing a fluctuating magnetic field. Here, we model the process using the central spin model's pure dephasing dynamics, where the interaction with the bath switches between two states at random. Further, the number $k$ of switches that take place within a time $t$ follows a Poisson distribution $P_k = \frac{(\gamma t)^k}{k!} e^{-\gamma t}$. Using these conditions the factor $\Lambda(t)$, as shown in the SM~\cite{supplemental}, becomes
\begin{align}\label{eq_Lambda}
    \Lambda(t) = e^{-\gamma t}\left[\cos\left(\mu \gamma t\right) + \frac{1}{\mu} \sin(\mu\gamma t)\right],
\end{align}
where $\mu = \sqrt{\frac{\epsilon^2 N}{\gamma^2} - 1}$. The system's evolution is non-Markovian for $\left(\frac{\epsilon \sqrt{N}}{\gamma}\right)^2 > 1$, otherwise it is Markovian~\cite{P_Kumar_2018}. 

We now find the canonical master equation for this RTN dynamics. The matrix elements of the dynamical map $\Phi_t$ for this model are $\phi_{11} = \phi_{44}= 1$ and $\phi_{22} = \Lambda(t)e^{-i\omega_0t}$; the rest of the elements $\phi_{jk}$ are zero. Correspondingly, the non-zero matrix elements of the Lindblad superoperator $\mathbb{L}_t$ are $l_{22} = \frac{\dot \Lambda(t) - i\omega_0\Lambda(t)}{\Lambda(t)}$ and $l_{22}^*$. The eigenvalues of the resulting Choi matrix are given by $\lambda_1 = -|l_{22}|$, $\lambda_2 = |l_{22}|$, and $\lambda_3 = \lambda_4 = 0$. The corresponding pseudo-Kraus operators, which are derived from the eigenvectors of the Choi matrix, are $E_1 = \frac{1}{\sqrt{2}}\begin{pmatrix} -l_{22}/|l_{22}|&0\\0&1 \end{pmatrix}$ and $E_2 = \frac{1}{\sqrt{2}}\begin{pmatrix} l_{22}/|l_{22}|&0\\0&1 \end{pmatrix}$. The canonical Hamiltonian for this model is thus given by $H^{\can} = \omega_0\sigma^z/2$. The Lindblad jump operators constituting the minimal dissipator for this model are $L_1 = -\left[(1 + l_{22}/|l_{22}|)/2\sqrt{2}\right]\sigma^z$ and $L_2 = \left[(-1 + l_{22}/|l_{22}|)/2\sqrt{2}\right]\sigma^z$. Consequently, the master equation is given by
\begin{align}\label{eq_RTN_master_equation}
    \frac{d\rho_S(t)}{dt} = -i\left[\frac{\omega_0}2\sigma^z, \rho_S(t)\right] - \frac{\dot\Lambda(t)}{2\Lambda(t)}\left[\sigma^z\rho_S(t)\sigma^z - \rho_S(t)\right].
\end{align}
This is a pure dephasing master equation. It is worth noting that even though a similar kind of master equation can be derived using the Hamiltonian $H = H_S + H_B + H_{SB} = \omega_0\sigma^z/2 + \sum_k\omega_k a_k^\dagger a_k + \sigma^z\sum_kg_k(a_k^\dagger + a_k)$ ($a_k$ being the bosonic annihilation operator for the mode $k$), the dephasing factor obtained from this Hamiltonian doesn't match the RTN dephasing factor (see~\cite{supplemental} for a proof). This brings out that a spin bath is required for a quantum mechanical treatment of the RTN, highlighting the ubiquity of the central spin model. 

Recently in~\cite{RTN_dissipative}, an open quantum system interacting with a non-Gaussian RTN bath was studied, demonstrating the system's dissipative dynamics with energy loss and gain terms, which is different from the pure dephasing dynamics shown here. This suggests that merging the dissipative and pure dephasing RTN system--bath interactions might result in a new phase covariant master equation. It would be interesting to investigate this further.  

\begin{figure*}
    \centering
    \includegraphics[width=1\linewidth]{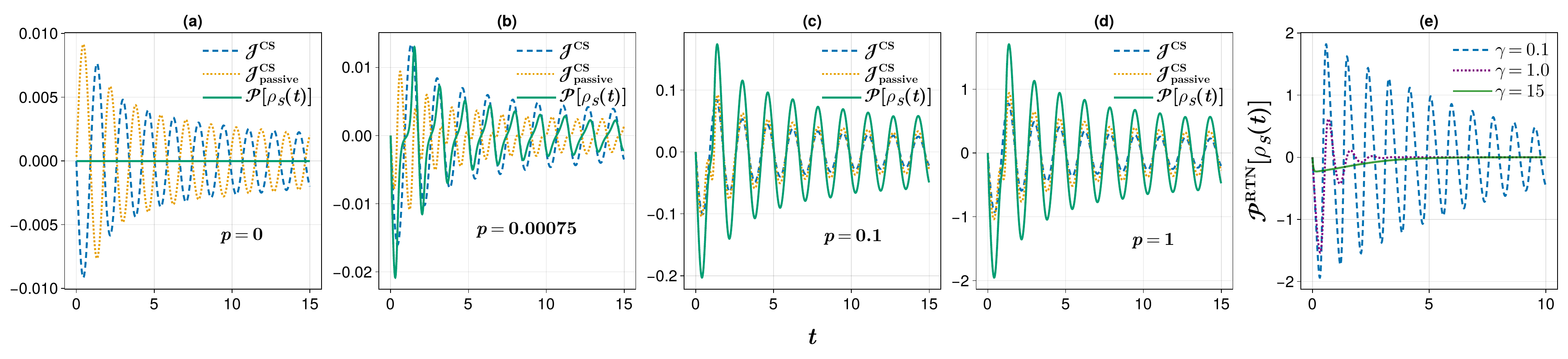}
    \caption{Variation of the heat current, $\mathcal{J}^\cs$, passive heat current $\mathcal{J}^\cs_{\rm passive}$, and charging power $\mathcal{P}[\rho_S(t)]$ for the dissipative central spin model, Eq.~\eqref{eq_CS_Ham} in (a), (b), (c), and (d) for different values of mixing probability $p$, and variation of the charging power $\mathcal{P}^\rtn[\rho_S(t)]$ with time for the RTN model in (e). In (a)-(d), the parameters are: $\omega = 1.0, \omega_0 = 1.5, N = 50, \epsilon = 0.5$, and $\beta = 0.5$. In (e), the parameters are: $\omega_0 = 1.5, N = 50, \epsilon = 0.5$, and the initial state is taken to be $\ket{\psi_S(0)} = \frac{\sqrt{3}}{2}\ket{0} + \frac{1}{2}\ket{1}$. }
    \label{fig_charging_power}
\end{figure*}
\section{Application: Quantum Thermodynamics}Having obtained the canonical master equations, we now demonstrate their utility by making an application to, for example, quantum thermodynamics. Consider the following form of first law of quantum thermodynamics $\frac{\der}{\der t}\Tr\left[H_S\rho_S(t)\right] = \Tr\left[\frac{\der}{\der t}H_S \rho_S(t)\right] + \Tr\left[H_S\frac{\der}{\der t}\rho_S(t)\right]$. The left side of this equation denotes the rate of the system's internal energy change. The first term on the right side of the equation is the rate of work done on the system, and the second term denotes the rate of the system's heat exchange, or heat current. Care should be taken here that in the computation of currents, for quantum thermodynamic consistency, $H_S$ is the bare system Hamiltonian and not the canonical Hamiltonian~\cite{Tiwari2025}. The heat current, for the central spin model's dissipative dynamics, is derived using Eq.~\eqref{eq_CS_master_eq}, and is readily shown to be  
\begin{align}
    \mathcal{J^\cs} = \omega_0\left[\dot \eta_t + (\dot \alpha_t - \dot \eta_t)\rho_{00}(0)\right].
\end{align}
A similar calculation for the RTN model gives $\mathcal{J^{\rm RTN}} = 0$. The complexity in calculating the heat current is significantly reduced when the master equation is known. For example, for the dissipative dynamics, it can be easily discerned from the form of the master equation~\eqref{eq_CS_master_eq} that only the energy gain and loss terms contribute to the heat current. Similarly, the structure of the RTN master equation~\eqref{eq_RTN_master_equation} reveals that its trace with the bare system Hamiltonian ($H_S = \frac{\omega_0}2\sigma^z$) would be zero. 

Further, the TLSs studied here can be modeled as a quantum battery~\cite{colloquium_QB, topological_QB, Fischer_QB}, with the bath as the charger, charging the battery via the system--bath interaction. To this end, the bound on the maximum work extraction is given by the ergotropy $\mathcal{W}[\rho_S(t)] = \Tr[H_S\rho_S(t)] - \Tr[\rho^pH_S]$, where $\rho^p$ is the passive state corresponding to the system~\cite{Allahverdyan_2004}. For a general TLS, with $H_S = \frac{\omega_0}{2}\sigma^z$ as the system Hamiltonian, the ergotropy is given by $\W[\rho_S(t)] = \frac{\omega_0}{2}\left[z_t + \sqrt{{x_t^2 + y_t^2 + z_t^2}}\right]$, where $k_t = \Tr[\sigma^k\rho_S(t)]$.  

The charging power of the quantum battery is defined as $\pow = \der\W/\der t$. A positive charging power indicates an increment in the battery's ergotropy, that is, charging of the battery, and correspondingly, a negative charging power indicates discharging. For the dissipative central spin model, it becomes $\pow[\rho_S(t)] = \mathcal{J^\cs} + \mathcal{J^\cs_{\rm passive}}$, where $\mathcal{J^\cs}$ is heat current derived above and $\mathcal{J^\cs_{\rm passive}} = \frac{\omega_0}{2}\left(\frac{x_t\dot x_t + y_t\dot y_t + z_t \dot z_t}{\sqrt{x_t^2 + y_t^2 + z_t^2}}\right)$ is the passive heat current, as it stems from the passive state. 
The charging power for the dissipative central spin model is depicted in Fig.~\ref{fig_charging_power}(a)-(d), where $\rho_S(0) = p\ket{\psi_S(0)}\bra{\psi_S(0)} + \frac{(1 - p)}{2}\mathbb{I}_S$ is considered as the system's initial state, with $\ket{\psi_S(0)} = \frac{\sqrt{3}}{2}\ket{0} + \frac{1}{2}\ket{1}$ and $p$ being the mixing probability. The value of $p$ decides whether the initial state is a maximally mixed or a pure state. It can be observed that for the maximally mixed state ($p = 0$), the charging power is zero as the heat current and the passive heat current, though non-zero, cancel each other. A very small increment in $p$ leads to a non-zero charging power, thereby highlighting the importance of coherence in charging the quantum battery. As we increase $p$, the oscillations of the heat current and the passive heat current synchronize, and thus, the amplitude of the charging power increases. It takes the highest values for the initial pure state. The oscillations in the charging power between the negative and positive values depict the discharging and charging of the battery, respectively. Further, the impact of the number of bath spins $N$ and interaction strength $\epsilon$ on the dynamics of the charging power and heat current is shown in SM~\cite{supplemental}.

Further, the charging power for the RTN model is exactly equal to the passive heat current, as $\mathcal{J}^\rtn = 0$.
The ergotropy and correspondingly the charging power for the RTN model are given by 
$
    \mathcal{W}^{\rtn}[\rho_S(t)] = \frac{\omega_0}{2}z(0)^\rtn+ \frac{\omega_0}{2}\sqrt{z(0)^\rtn + 4\left|\rho_S^{01}(0)\right|^2\Lambda(t)^2}
$, and 
\begin{align}
    \mathcal{P}^{\rtn}[\rho_S(t)] = \frac{2\omega_0\left|\rho_S^{01}(0)\right|^2\Lambda(t)\dot\Lambda(t)}{\sqrt{\left[2\rho_S^{00}(0) - 1\right]^2 + 4\left|\rho_S^{01}(0)\right|^2\Lambda(t)^2}},
\end{align}
respectively, where $z(0)^\rtn = 2\rho_S^{00}(0) - 1$.
The charging power for the RTN model is non-zero when the state has a non-zero coherence. Further, the incoherent part of the ergotropy for this model doesn't change, highlighting an advantage in the battery's charge storage. 
Figure~\ref{fig_charging_power}(e) illustrates the charging power of the RTN model. It is evident that a reduction in the switching rate $\gamma$ results in increased oscillation amplitudes of the charging power, signifying an enhanced charge and discharge rate. When $\left(\epsilon\sqrt{N}/\gamma\right)^2< 1$, the charging power remains negative and changes monotonically, as depicted by the green curve, indicating an absence of charging within this Markovian regime, Eq.~\eqref{eq_Lambda}.  

We note here that the central spin model's dissipative dynamics has also been developed in~\cite{Tiwari2025}. Here, we clearly outline the novel aspects achieved in this work. (a). \textit{General method (pseudo-Kraus operators):} The recipe introduced here, that is, deriving the master equation systematically from the dynamical map via the Choi matrix spectral decomposition and pseudo-Kraus operators, is a methodological advance developed in this work. While in~\cite{Tiwari2025}, the minimal dissipation is applied directly, the present formulation provides a structured, step-by-step algorithm applicable to any CPTP dynamical map of a two-level system, including maps obtained numerically (e.g., via HEOM or reaction-coordinate mapping).
(b). \textit{Identification of exact phase covariance:} While the dynamical map was obtained in~\cite{Tiwari2025}, the identification that the resulting master equation is exactly phase covariant, which is also valid in the strong-coupling regime and for a finite bath at arbitrary temperature, has been established in this work. This structural identification, together with its physical implications, is new.
(c). \textit{RTN model and bosonic incompatibility:} The stochastic pure-dephasing interaction, the microscopic quantum derivation of the RTN channel, and the proof that a bosonic bath cannot replicate the RTN dephasing factor are entirely new results. Furthermore, the analysis of heat currents, charging power, and the direct relationship between quantum heat current and charging power for both models is new.

\section{Conclusions}In this work, a novel technique to obtain a master equation from the dynamics of a quantum system using the concept of minimal dissipator was developed, and the Lindbladian and dynamical map of a general two-level system were explored. This was applied to an important class of finite bath models—the central spin model. An exact dynamical map of the central spin model for two different types of system-bath interactions, dissipative and pure dephasing interactions, was developed, leading to novel quantum channels and corresponding Kraus operators. The dissipative interaction led to a new phase covariant master equation, which was exact even in the strong-coupling regime. This analysis lays the groundwork for a foray into channel capacity, information, and quantum thermodynamics, among others, in finite-bath open quantum systems. The pure dephasing interaction with stochastic random coupling produced an RTN evolution, paving the way for the study of microscopic stochastic system-bath coupling models. This type of coupling would be conducive for studying quantum chaos in a wide range of open quantum systems; thus, for example, the impact of random stochastic atom-field coupling on the superradiant phase transition and quantum chaos can be investigated. Furthermore, the application of the master equations obtained was demonstrated in quantum thermodynamics, where a relationship between the charging power and the quantum heat current of the TLSs was shown. Owing to the wide experimental handle of the central spin model using Rydberg atoms~\cite{Tomza_cs_expt_1}, Nitrogen-Vacancy center~\cite{Cirac_expt_cs_2, Hanson_cs_expt_4}, and NMR spin systems~\cite{TSMahesh_2022}, the applications borne out of the present study can be envisaged experimentally. Thus, for example, the NMR spin systems~\cite{TSMahesh_2022}, which have been used for an experimental quantum battery using numerous molecular structures, can be tuned to study the heat currents and charging power experimentally in the central spin model. Further, the random coupling in the NMR spin systems can be used to experimentally realize the RTN dynamics. Using the Kraus operators developed here, and the above experimental techniques, the Fisher information and thus quantum metrology applications can be performed practically. Ultimately, this work establishes a rigorous foundation for finite-bath dynamics, bridging fundamental theory with applications.

\section{Acknowledgments}
S.B. acknowledges useful discussions with T.S. Mahesh.  

\bibliography{reference}

@article{Hayden_2022,
doi = {10.1088/1751-8121/ac65c2},
url = {https://dx.doi.org/10.1088/1751-8121/ac65c2},
year = {2022},
month = {may},
publisher = {IOP Publishing},
volume = {55},
number = {22},
pages = {225302},
author = {Patrick Hayden and Jonathan Sorce},
title = {A canonical Hamiltonian for open quantum systems},
journal = {Journal of Physics A: Mathematical and Theoretical}
}

@article{breuer_effective_Ham1,
  title = {Open-system approach to nonequilibrium quantum thermodynamics at arbitrary coupling},
  author = {Colla, Alessandra and Breuer, Heinz-Peter},
  journal = {Phys. Rev. A},
  volume = {105},
  issue = {5},
  pages = {052216},
  numpages = {8},
  year = {2022},
  month = {May},
  publisher = {American Physical Society},
  doi = {10.1103/PhysRevA.105.052216},
  url = {https://link.aps.org/doi/10.1103/PhysRevA.105.052216}
}

@book{BreurPoqs,
    author = {Breuer, Heinz-Peter and Petruccione, Francesco},
    title = "{The Theory of Open Quantum Systems}",
    publisher = {Oxford University Press},
    year = {2007},
    month = {01},
    isbn = {9780199213900},
    doi = {10.1093/acprof:oso/9780199213900.001.0001},
    url = {https://doi.org/10.1093/acprof:oso/9780199213900.001.0001},
}

@book{banerjee2018open,
  title={Open Quantum Systems: Dynamics of Nonclassical Evolution},
  author={Banerjee, S.},
  isbn={9789811331824},
  series={Texts and Readings in Physical Sciences},
  url={https://books.google.co.in/books?id=s-11DwAAQBAJ},
  year={2018},
  publisher={Springer Nature Singapore}
}

@article{Breuer_colloq_nonM,
  title = {Colloquium: Non-Markovian dynamics in open quantum systems},
  author = {Breuer, Heinz-Peter and Laine, Elsi-Mari and Piilo, Jyrki and Vacchini, Bassano},
  journal = {Rev. Mod. Phys.},
  volume = {88},
  issue = {2},
  pages = {021002},
  numpages = {24},
  year = {2016},
  month = {Apr},
  publisher = {American Physical Society},
  doi = {10.1103/RevModPhys.88.021002},
  url = {https://link.aps.org/doi/10.1103/RevModPhys.88.021002}
}

@article{GKS_paper,
    author = {Gorini, Vittorio and Kossakowski, Andrzej and Sudarshan, E. C. G.},
    title = "{Completely positive dynamical semigroups of N‐level systems}",
    journal = {Journal of Mathematical Physics},
    volume = {17},
    number = {5},
    pages = {821-825},
    year = {1976},
    month = {05},
    issn = {0022-2488},
    doi = {10.1063/1.522979},
    url = {https://doi.org/10.1063/1.522979},
    eprint = {https://pubs.aip.org/aip/jmp/article-pdf/17/5/821/19090720/821\_1\_online.pdf},
}

@article{lindblad_1976,
author={Lindblad, G.},
title={On the generators of quantum dynamical semigroups},
journal={Communications in Mathematical Physics},
year={1976},
month={Jun},
day={01},
volume={48},
number={2},
pages={119-130},
issn={1432-0916},
doi={10.1007/BF01608499},
url={https://doi.org/10.1007/BF01608499}
}

@article{Chiranjib_2017,
  title = {Dynamics and thermodynamics of a central spin immersed in a spin bath},
  author = {Mukhopadhyay, Chiranjib and Bhattacharya, Samyadeb and Misra, Avijit and Pati, Arun Kumar},
  journal = {Phys. Rev. A},
  volume = {96},
  issue = {5},
  pages = {052125},
  numpages = {13},
  year = {2017},
  month = {Nov},
  publisher = {American Physical Society},
  doi = {10.1103/PhysRevA.96.052125},
  url = {https://link.aps.org/doi/10.1103/PhysRevA.96.052125}
}

@article{landi_entropy_production,
  title = {Irreversible entropy production: From classical to quantum},
  author = {Landi, Gabriel T. and Paternostro, Mauro},
  journal = {Rev. Mod. Phys.},
  volume = {93},
  issue = {3},
  pages = {035008},
  numpages = {58},
  year = {2021},
  month = {Sep},
  publisher = {American Physical Society},
  doi = {10.1103/RevModPhys.93.035008},
  url = {https://link.aps.org/doi/10.1103/RevModPhys.93.035008}
}

@Article{kosloff_2022,
AUTHOR = {Kosloff, Ronnie},
TITLE = {Quantum Thermodynamics: A Dynamical Viewpoint},
JOURNAL = {Entropy},
VOLUME = {15},
YEAR = {2013},
NUMBER = {6},
PAGES = {2100--2128},
URL = {https://www.mdpi.com/1099-4300/15/6/2100},
ISSN = {1099-4300},
DOI = {10.3390/e15062100}
}

@book{deffner_book,
author = {Deffner, Sebastian and Campbell, Steve},
title = {Quantum Thermodynamics},
publisher = {Morgan \& Claypool Publishers},
year = {2019},
series = {2053-2571},
isbn = {978-1-64327-658-8},
url = {https://dx.doi.org/10.1088/2053-2571/ab21c6},
doi = {10.1088/2053-2571/ab21c6}
}

@article{vega_alonso,
  title = {Dynamics of non-Markovian open quantum systems},
  author = {de Vega, In\'es and Alonso, Daniel},
  journal = {Rev. Mod. Phys.},
  volume = {89},
  issue = {1},
  pages = {015001},
  numpages = {58},
  year = {2017},
  month = {Jan},
  publisher = {American Physical Society},
  doi = {10.1103/RevModPhys.89.015001},
  url = {https://link.aps.org/doi/10.1103/RevModPhys.89.015001}
}

@article{Filippov2020,
  title = {Phase Covariant Qubit Dynamics and Divisibility},
  volume = {41},
  ISSN = {1818-9962},
  url = {http://dx.doi.org/10.1134/S1995080220040095},
  DOI = {10.1134/s1995080220040095},
  number = {4},
  journal = {Lobachevskii Journal of Mathematics},
  publisher = {Pleiades Publishing Ltd},
  author = {Filippov,  S. N. and Glinov,  A. N. and Lepp\"{a}j\"{a}rvi,  L.},
  year = {2020},
  month = apr,
  pages = {617–630}
}

@article{prokofev_stamp_2000,
doi = {10.1088/0034-4885/63/4/204},
url = {https://dx.doi.org/10.1088/0034-4885/63/4/204},
year = {2000},
month = {apr},
publisher = {},
volume = {63},
number = {4},
pages = {669},
author = {N V Prokof'ev and  P C E Stamp},
title = {Theory 
of the spin bath},
journal = {Reports on Progress in Physics}
}

@article{Devvrat_central_spin1,
  title = {Dynamics of two central spins immersed in spin baths},
  author = {Tiwari, Devvrat and Datta, Shounak and Bhattacharya, Samyadeb and Banerjee, Subhashish},
  journal = {Phys. Rev. A},
  volume = {106},
  issue = {3},
  pages = {032435},
  numpages = {15},
  year = {2022},
  month = {Sep},
  publisher = {American Physical Society},
  doi = {10.1103/PhysRevA.106.032435},
  url = {https://link.aps.org/doi/10.1103/PhysRevA.106.032435}
}

@article{Allahverdyan_2004,
doi = {10.1209/epl/i2004-10101-2},
url = {https://dx.doi.org/10.1209/epl/i2004-10101-2},
year = {2004},
month = {aug},
publisher = {},
volume = {67},
number = {4},
pages = {565},
author = {A. E. Allahverdyan and  R. Balian and  Th. M. Nieuwenhuizen},
title = {Maximal work extraction from finite quantum systems},
journal = {Europhysics Letters}
}

@article{Hughes_reaction_coordinate,
     author = {Hughes, Keith H. and Christ, Clara D. and Burghardt, Irene},
    title = {Effective-mode representation of non-Markovian dynamics: A hierarchical approximation of the spectral density. I. Application to single surface dynamics},
    journal = {The Journal of Chemical Physics},
    volume = {131},
    number = {2},
    pages = {024109},
    year = {2009},
    month = {07},
    issn = {0021-9606},
    doi = {10.1063/1.3159671},
    url = {https://doi.org/10.1063/1.3159671}
}

@article{TSMahesh_2022,
  title = {Experimental investigation of a quantum battery using star-topology NMR spin systems},
  author = {Joshi, Jitendra and Mahesh, T. S.},
  journal = {Phys. Rev. A},
  volume = {106},
  issue = {4},
  pages = {042601},
  numpages = {8},
  year = {2022},
  month = {Oct},
  publisher = {American Physical Society},
  doi = {10.1103/PhysRevA.106.042601},
  url = {https://link.aps.org/doi/10.1103/PhysRevA.106.042601}
}

@article{Tiwari2025,
  title = {Strong coupling non-Markovian quantum thermodynamics of a finite-bath system},
  volume = {162},
  ISSN = {1089-7690},
  url = {http://dx.doi.org/10.1063/5.0254029},
  DOI = {10.1063/5.0254029},
  number = {11},
  journal = {The Journal of Chemical Physics},
  publisher = {AIP Publishing},
  author = {Tiwari,  Devvrat and Bose,  Baibhab and Banerjee,  Subhashish},
  year = {2025},
  month = mar 
}

@article{KRAUS1971,
title = {General state changes in quantum theory},
journal = {Annals of Physics},
volume = {64},
number = {2},
pages = {311-335},
year = {1971},
doi = {https://doi.org/10.1016/0003-4916(71)90108-4},
url = {https://www.sciencedirect.com/science/article/pii/0003491671901084},
author = {K Kraus}
}

@article{Haase_2018,
doi = {10.1088/1367-2630/aab67f},
url = {https://dx.doi.org/10.1088/1367-2630/aab67f},
year = {2018},
month = {may},
publisher = {IOP Publishing},
volume = {20},
number = {5},
pages = {053009},
author = {Haase, J F and Smirne, A and Kołodyński, J and Demkowicz-Dobrzański, R and Huelga, S F},
title = {Fundamental limits to frequency estimation: a comprehensive microscopic perspective},
journal = {New Journal of Physics}
}

@article{Smirne_2016,
  title = {Ultimate Precision Limits for Noisy Frequency Estimation},
  author = {Smirne, Andrea and Ko\l{}ody\ifmmode \acute{n}\else \'{n}\fi{}ski, Jan and Huelga, Susana F. and Demkowicz-Dobrza\ifmmode \acute{n}\else \'{n}\fi{}ski, Rafa\l{}},
  journal = {Phys. Rev. Lett.},
  volume = {116},
  issue = {12},
  pages = {120801},
  numpages = {6},
  year = {2016},
  month = {Mar},
  publisher = {American Physical Society},
  doi = {10.1103/PhysRevLett.116.120801},
  url = {https://link.aps.org/doi/10.1103/PhysRevLett.116.120801}
}

@article{Samyadeb_2017,
  title = {Exact master equation for a spin interacting with a spin bath: Non-Markovianity and negative entropy production rate},
  author = {Bhattacharya, Samyadeb and Misra, Avijit and Mukhopadhyay, Chiranjib and Pati, Arun Kumar},
  journal = {Phys. Rev. A},
  volume = {95},
  issue = {1},
  pages = {012122},
  numpages = {10},
  year = {2017},
  month = {Jan},
  publisher = {American Physical Society},
  doi = {10.1103/PhysRevA.95.012122},
  url = {https://link.aps.org/doi/10.1103/PhysRevA.95.012122}
}

@article{Holstein-Primakoff_paper,
  title = {Field Dependence of the Intrinsic Domain Magnetization of a Ferromagnet},
  author = {Holstein, T. and Primakoff, H.},
  journal = {Phys. Rev.},
  volume = {58},
  issue = {12},
  pages = {1098--1113},
  numpages = {0},
  year = {1940},
  month = {Dec},
  publisher = {American Physical Society},
  doi = {10.1103/PhysRev.58.1098},
  url = {https://link.aps.org/doi/10.1103/PhysRev.58.1098}
}

@Article{Utagi2020,
author={Utagi, Shrikant
and Srikanth, R.
and Banerjee, Subhashish},
title={Temporal self-similarity of quantum dynamical maps as a concept of memorylessness},
journal={Scientific Reports},
year={2020},
month={Sep},
day={14},
volume={10},
number={1},
pages={15049},
issn={2045-2322},
doi={10.1038/s41598-020-72211-3},
url={https://doi.org/10.1038/s41598-020-72211-3}
}

@article{rtn_model_1,
  title = {Depolarizing channel as a completely positive map with memory},
  author = {Daffer, Sonja and W\'odkiewicz, Krzysztof and Cresser, James D. and McIver, John K.},
  journal = {Phys. Rev. A},
  volume = {70},
  issue = {1},
  pages = {010304},
  numpages = {4},
  year = {2004},
  month = {Jul},
  publisher = {American Physical Society},
  doi = {10.1103/PhysRevA.70.010304},
  url = {https://link.aps.org/doi/10.1103/PhysRevA.70.010304}
}

@article{colla2025_perturbation,
  title = {Unveiling coherent dynamics in non-Markovian open quantum systems: Exact expression and recursive perturbation expansion},
  author = {Colla, Alessandra and Breuer, Heinz-Peter and Gasbarri, Giulio},
  journal = {Phys. Rev. A},
  volume = {112},
  issue = {5},
  pages = {L050203},
  numpages = {7},
  year = {2025},
  month = {Nov},
  publisher = {American Physical Society},
  doi = {10.1103/n5nl-gn1y},
  url = {https://link.aps.org/doi/10.1103/n5nl-gn1y}
}

@article{RTN_dissipative,
  title = {Dynamics of a Quantum System Interacting with White Non-Gaussian Baths: Poisson Noise Master Equation},
  author = {Funo, Ken and Ishizaki, Akihito},
  journal = {Phys. Rev. Lett.},
  volume = {132},
  issue = {17},
  pages = {170402},
  numpages = {7},
  year = {2024},
  month = {Apr},
  publisher = {American Physical Society},
  doi = {10.1103/PhysRevLett.132.170402},
  url = {https://link.aps.org/doi/10.1103/PhysRevLett.132.170402}
}

@article{RTN1_Bergli_2009,
doi = {10.1088/1367-2630/11/2/025002},
url = {https://dx.doi.org/10.1088/1367-2630/11/2/025002},
year = {2009},
month = {feb},
publisher = {},
volume = {11},
number = {2},
pages = {025002},
author = {Bergli, J and Galperin, Y M and Altshuler, B L},
title = {Decoherence in qubits due to low-frequency noise},
journal = {New Journal of Physics}
}

@Article{RTN3_Cai2020,
author={Cai, Xiangji},
title={Quantum dephasing induced by non-Markovian random telegraph noise},
journal={Scientific Reports},
year={2020},
month={Jan},
day={09},
volume={10},
number={1},
pages={88},
issn={2045-2322},
doi={10.1038/s41598-019-57081-8},
url={https://doi.org/10.1038/s41598-019-57081-8}
}

@Article{Colla2025-expt,
author={Colla, Alessandra
and Hasse, Florian
and Palani, Deviprasath
and Schaetz, Tobias
and Breuer, Heinz-Peter
and Warring, Ulrich},
title={Observing time-dependent energy level renormalisation in an ultrastrongly coupled open system},
journal={Nature Communications},
year={2025},
month={Mar},
day={13},
volume={16},
number={1},
pages={2502},
issn={2041-1723},
doi={10.1038/s41467-025-57840-4},
url={https://doi.org/10.1038/s41467-025-57840-4}
}

@misc{supplemental,
    title = {The Supplemental Material discusses the theory of the minimal dissipation briefly, provides details on a general dynamical map of a two-level system, the exact solution of the central spin model, and the random telegraph noise master equation, and is attached below.} 
}

@article{Vacchini_2010,
  title = {Nakajima-Zwanzig versus time-convolutionless master equation for the non-Markovian dynamics of a two-level system},
  author = {Smirne, Andrea and Vacchini, Bassano},
  journal = {Phys. Rev. A},
  volume = {82},
  issue = {2},
  pages = {022110},
  numpages = {10},
  year = {2010},
  month = {Aug},
  publisher = {American Physical Society},
  doi = {10.1103/PhysRevA.82.022110},
  url = {https://link.aps.org/doi/10.1103/PhysRevA.82.022110}
}

@ARTICLE{SO_rice,
  author={Rice, S. O.},
  journal={The Bell System Technical Journal}, 
  title={Mathematical analysis of random noise}, 
  year={1944},
  volume={23},
  number={3},
  pages={282-332},
  doi={10.1002/j.1538-7305.1944.tb00874.x}
}

@article{Nakamura_2002,
  title = {Charge Echo in a Cooper-Pair Box},
  author = {Nakamura, Y. and Pashkin, Yu. A. and Yamamoto, T. and Tsai, J. S.},
  journal = {Phys. Rev. Lett.},
  volume = {88},
  issue = {4},
  pages = {047901},
  numpages = {4},
  year = {2002},
  month = {Jan},
  publisher = {American Physical Society},
  doi = {10.1103/PhysRevLett.88.047901},
  url = {https://link.aps.org/doi/10.1103/PhysRevLett.88.047901}
}

@article{Zorin_1996,
  title = {Background charge noise in metallic single-electron tunneling devices},
  author = {Zorin, A. B. and Ahlers, F.-J. and Niemeyer, J. and Weimann, T. and Wolf, H. and Krupenin, V. A. and Lotkhov, S. V.},
  journal = {Phys. Rev. B},
  volume = {53},
  issue = {20},
  pages = {13682--13687},
  numpages = {0},
  year = {1996},
  month = {May},
  publisher = {American Physical Society},
  doi = {10.1103/PhysRevB.53.13682},
  url = {https://link.aps.org/doi/10.1103/PhysRevB.53.13682}
}

@article{P_Kumar_2018,
author = {Kumar, N. Pradeep and Banerjee, Subhashish and Srikanth, R. and Jagadish, Vinayak and Petruccione, Francesco},
title = {Non-Markovian Evolution: a Quantum Walk Perspective},
journal = {Open Systems \& Information Dynamics},
volume = {25},
number = {03},
pages = {1850014},
year = {2018},
doi = {10.1142/S1230161218500142},
URL = {https://doi.org/10.1142/S1230161218500142}
}

@article{Tomza_cs_expt_1,
  title = {Quantum simulation of the central spin model with a Rydberg atom and polar molecules in optical tweezers},
  author = {Dobrzyniecki, Jacek and Tomza, Micha\l{}},
  journal = {Phys. Rev. A},
  volume = {108},
  issue = {5},
  pages = {052618},
  numpages = {23},
  year = {2023},
  month = {Nov},
  publisher = {American Physical Society},
  doi = {10.1103/PhysRevA.108.052618},
  url = {https://link.aps.org/doi/10.1103/PhysRevA.108.052618}
}

@article{Cirac_expt_cs_2,
  title = {Optical Superradiance from Nuclear Spin Environment of Single-Photon Emitters},
  author = {Kessler, E. M. and Yelin, S. and Lukin, M. D. and Cirac, J. I. and Giedke, G.},
  journal = {Phys. Rev. Lett.},
  volume = {104},
  issue = {14},
  pages = {143601},
  numpages = {4},
  year = {2010},
  month = {Apr},
  publisher = {American Physical Society},
  doi = {10.1103/PhysRevLett.104.143601},
  url = {https://link.aps.org/doi/10.1103/PhysRevLett.104.143601}
}

@article{Hanson_cs_expt_4,
author = {R. Hanson  and V. V. Dobrovitski  and A. E. Feiguin  and O. Gywat  and D. D. Awschalom },
title = {Coherent Dynamics of a Single Spin Interacting with an Adjustable Spin Bath},
journal = {Science},
volume = {320},
number = {5874},
pages = {352-355},
year = {2008},
doi = {10.1126/science.1155400},
URL = {https://www.science.org/doi/abs/10.1126/science.1155400},
eprint = {https://www.science.org/doi/pdf/10.1126/science.1155400}
}

@BOOK{Louisell1973,
  title     = "Quantum statistical properties of radiation",
  author    = "Louisell, William H",
  publisher = "John Wiley \& Sons",
  series    = "Pure \& Applied Optics S.",
  month     =  jul,
  year      =  1973,
  address   = "Nashville, TN",
  language  = "en"
}

@book{Weiss2011,
  title = {Quantum Dissipative Systems},
  ISBN = {9789814374927},
  url = {http://dx.doi.org/10.1142/8334},
  DOI = {10.1142/8334},
  publisher = {WORLD SCIENTIFIC},
  author = {Weiss,  Ulrich},
  year = {2011},
  month = nov 
}

@article{Hall_andersson_2014,
  title = {Canonical form of master equations and characterization of non-Markovianity},
  author = {Hall, Michael J. W. and Cresser, James D. and Li, Li and Andersson, Erika},
  journal = {Phys. Rev. A},
  volume = {89},
  issue = {4},
  pages = {042120},
  numpages = {11},
  year = {2014},
  month = {Apr},
  publisher = {American Physical Society},
  doi = {10.1103/PhysRevA.89.042120},
  url = {https://link.aps.org/doi/10.1103/PhysRevA.89.042120}
}

@article{Rivas_2014,
doi = {10.1088/0034-4885/77/9/094001},
url = {https://dx.doi.org/10.1088/0034-4885/77/9/094001},
year = {2014},
month = {aug},
publisher = {IOP Publishing},
volume = {77},
number = {9},
pages = {094001},
author = {Rivas, \'Angel and Huelga, Susana F and Plenio, Martin B},
title = {Quantum non-Markovianity: characterization, quantification and detection},
journal = {Reports on Progress in Physics}
}

@article{CHRUSCINSKI20221,
title = {Dynamical maps beyond Markovian regime},
journal = {Physics Reports},
volume = {992},
pages = {1-85},
year = {2022},
issn = {0370-1573},
doi = {https://doi.org/10.1016/j.physrep.2022.09.003},
url = {https://www.sciencedirect.com/science/article/pii/S0370157322003428},
author = {Dariusz Chruściński}
}

@article{Landi_heat_current_RMP,
  title = {Nonequilibrium boundary-driven quantum systems: Models, methods, and properties},
  author = {Landi, Gabriel T. and Poletti, Dario and Schaller, Gernot},
  journal = {Rev. Mod. Phys.},
  volume = {94},
  issue = {4},
  pages = {045006},
  numpages = {58},
  year = {2022},
  month = {Dec},
  publisher = {American Physical Society},
  doi = {10.1103/RevModPhys.94.045006},
  url = {https://link.aps.org/doi/10.1103/RevModPhys.94.045006}
}

@article{Landi_current_paper_2,
  title = {Current Fluctuations in Open Quantum Systems: Bridging the Gap Between Quantum Continuous Measurements and Full Counting Statistics},
  author = {Landi, Gabriel T. and Kewming, Michael J. and Mitchison, Mark T. and Potts, Patrick P.},
  journal = {PRX Quantum},
  volume = {5},
  issue = {2},
  pages = {020201},
  numpages = {86},
  year = {2024},
  month = {Apr},
  publisher = {American Physical Society},
  doi = {10.1103/PRXQuantum.5.020201},
  url = {https://link.aps.org/doi/10.1103/PRXQuantum.5.020201}
}

@article{devvrat_thermal_circuit_paper,
  title = {Quantum Thermal Analogs of Electric Circuits: A Universal Approach},
  author = {Tiwari, Devvrat and Bhattacharya, Samyadeb and Banerjee, Subhashish},
  journal = {Phys. Rev. Lett.},
  volume = {135},
  issue = {2},
  pages = {020404},
  numpages = {6},
  year = {2025},
  month = {Jul},
  publisher = {American Physical Society},
  doi = {10.1103/5x8m-bhgd},
  url = {https://link.aps.org/doi/10.1103/5x8m-bhgd}
}

@article{Plenio_open_QSL,
  title = {Quantum Speed Limits in Open System Dynamics},
  author = {del Campo, A. and Egusquiza, I. L. and Plenio, M. B. and Huelga, S. F.},
  journal = {Phys. Rev. Lett.},
  volume = {110},
  issue = {5},
  pages = {050403},
  numpages = {5},
  year = {2013},
  month = {Jan},
  publisher = {American Physical Society},
  doi = {10.1103/PhysRevLett.110.050403},
  url = {https://link.aps.org/doi/10.1103/PhysRevLett.110.050403}
}

@article{Deffner_2017,
doi = {10.1088/1751-8121/aa86c6},
url = {https://doi.org/10.1088/1751-8121/aa86c6},
year = {2017},
month = {oct},
publisher = {IOP Publishing},
volume = {50},
number = {45},
pages = {453001},
author = {Deffner, Sebastian and Campbell, Steve},
title = {Quantum speed limits: from Heisenberg’s uncertainty principle to optimal quantum control},
journal = {Journal of Physics A: Mathematical and Theoretical}
}

@BOOK{Agarwal2012,
  title     = "Quantum optics",
  author    = "Agarwal, Girish S",
  publisher = "Cambridge University Press",
  month     =  nov,
  year      =  2012,
  address   = "Cambridge, England",
  language  = "en"
}

@book{Scully_Zubairy_1997, 
place={Cambridge}, 
title={Quantum Optics}, 
publisher={Cambridge University Press}, 
author={Scully, Marlan O. and Zubairy, M. Suhail}, 
year={1997}
}

@article{Finite_bath_esposito,
  title = {Quantum master equation for a system influencing its environment},
  author = {Esposito, Massimiliano and Gaspard, Pierre},
  journal = {Phys. Rev. E},
  volume = {68},
  issue = {6},
  pages = {066112},
  numpages = {18},
  year = {2003},
  month = {Dec},
  publisher = {American Physical Society},
  doi = {10.1103/PhysRevE.68.066112},
  url = {https://link.aps.org/doi/10.1103/PhysRevE.68.066112}
}

@article{Finite_bath_Strasberg_1,
  title = {Open quantum systems coupled to finite baths: A hierarchy of master equations},
  author = {Riera-Campeny, Andreu and Sanpera, Anna and Strasberg, Philipp},
  journal = {Phys. Rev. E},
  volume = {105},
  issue = {5},
  pages = {054119},
  numpages = {17},
  year = {2022},
  month = {May},
  publisher = {American Physical Society},
  doi = {10.1103/PhysRevE.105.054119},
  url = {https://link.aps.org/doi/10.1103/PhysRevE.105.054119}
}

@article{Mahler_2011,
  title = {Operational approach to fluctuations of thermodynamic variables in finite quantum systems},
  author = {Jahnke, T. and Lan\'ery, S. and Mahler, G.},
  journal = {Phys. Rev. E},
  volume = {83},
  issue = {1},
  pages = {011109},
  numpages = {5},
  year = {2011},
  month = {Jan},
  publisher = {American Physical Society},
  doi = {10.1103/PhysRevE.83.011109},
  url = {https://link.aps.org/doi/10.1103/PhysRevE.83.011109}
}

@Article{Pekola2016,
author={Pekola, J. P.
and Suomela, S.
and Galperin, Y. M.},
title={Finite-Size Bath in Qubit Thermodynamics},
journal={Journal of Low Temperature Physics},
year={2016},
month={Sep},
day={01},
volume={184},
number={5},
pages={1015-1029},
issn={1573-7357},
doi={10.1007/s10909-016-1618-5},
url={https://doi.org/10.1007/s10909-016-1618-5}
}

@article{Pekola_finite_bath_2,
author = {Pekola, Jukka P. and Karimi, Bayan and Cattaneo, Marco and Maniscalco, Sabrina},
title = {Long-Time Relaxation of a Finite Spin Bath Linearly Coupled to a Qubit},
journal = {Open Systems \& Information Dynamics},
volume = {30},
number = {02},
pages = {2350009},
year = {2023},
doi = {10.1142/S1230161223500099},
URL = {https://doi.org/10.1142/S1230161223500099}
}

@article{finite_bath_Sanpera,
  title = {Quantum Systems Correlated with a Finite Bath: Nonequilibrium Dynamics and Thermodynamics},
  author = {Riera-Campeny, Andreu and Sanpera, Anna and Strasberg, Philipp},
  journal = {PRX Quantum},
  volume = {2},
  issue = {1},
  pages = {010340},
  numpages = {24},
  year = {2021},
  month = {Mar},
  publisher = {American Physical Society},
  doi = {10.1103/PRXQuantum.2.010340},
  url = {https://link.aps.org/doi/10.1103/PRXQuantum.2.010340}
}

@article{Arenz_2014,
doi = {10.1088/1367-2630/16/6/065023},
url = {https://doi.org/10.1088/1367-2630/16/6/065023},
year = {2014},
month = {jun},
publisher = {IOP Publishing},
volume = {16},
number = {6},
pages = {065023},
author = {Arenz, Christian and Gualdi, Giulia and Burgarth, Daniel},
title = {Control of open quantum systems: case study of the central spin model},
journal = {New Journal of Physics}
}

@misc{biswas2025,
      title={The Floquet central spin model: A platform to realize eternal time crystals, entanglement steering, and multiparameter metrology}, 
      author={Hillol Biswas and Sayan Choudhury},
      year={2025},
      eprint={2501.18472},
      archivePrefix={arXiv},
      primaryClass={quant-ph},
      url={https://arxiv.org/abs/2501.18472}, 
}

@article{central_spin_quantum_dot1,
  title = {Thermal electron spin flip in quantum dots},
  author = {Fuchs, Moritz and Krau\ss{}, Felix and Hetterich, Daniel and Trauzettel, Bj\"orn},
  journal = {Phys. Rev. B},
  volume = {92},
  issue = {3},
  pages = {035310},
  numpages = {8},
  year = {2015},
  month = {Jul},
  publisher = {American Physical Society},
  doi = {10.1103/PhysRevB.92.035310},
  url = {https://link.aps.org/doi/10.1103/PhysRevB.92.035310}
}

@article{Campbell_2023,
  title = {Thermometry of strongly correlated fermionic quantum systems using impurity probes},
  author = {Mihailescu, George and Campbell, Steve and Mitchell, Andrew K.},
  journal = {Phys. Rev. A},
  volume = {107},
  issue = {4},
  pages = {042614},
  numpages = {16},
  year = {2023},
  month = {Apr},
  publisher = {American Physical Society},
  doi = {10.1103/PhysRevA.107.042614},
  url = {https://link.aps.org/doi/10.1103/PhysRevA.107.042614}
}

@article{Paladino_2002,
  title = {Decoherence and $1/\mathit{f}$ Noise in Josephson Qubits},
  author = {Paladino, E. and Faoro, L. and Falci, G. and Fazio, Rosario},
  journal = {Phys. Rev. Lett.},
  volume = {88},
  issue = {22},
  pages = {228304},
  numpages = {4},
  year = {2002},
  month = {May},
  publisher = {American Physical Society},
  doi = {10.1103/PhysRevLett.88.228304},
  url = {https://link.aps.org/doi/10.1103/PhysRevLett.88.228304}
}

@article{Bergli_2006_2,
  title = {Non-Gaussian Low-Frequency Noise as a Source of Qubit Decoherence},
  author = {Galperin, Y. M. and Altshuler, B. L. and Bergli, J. and Shantsev, D. V.},
  journal = {Phys. Rev. Lett.},
  volume = {96},
  issue = {9},
  pages = {097009},
  numpages = {4},
  year = {2006},
  month = {Mar},
  publisher = {American Physical Society},
  doi = {10.1103/PhysRevLett.96.097009},
  url = {https://link.aps.org/doi/10.1103/PhysRevLett.96.097009}
}

@article{Joynt_2008,
  title = {Transfer matrix solution of a model of qubit decoherence due to telegraph noise},
  author = {Cheng, Bin and Wang, Qiang-Hua and Joynt, Robert},
  journal = {Phys. Rev. A},
  volume = {78},
  issue = {2},
  pages = {022313},
  numpages = {7},
  year = {2008},
  month = {Aug},
  publisher = {American Physical Society},
  doi = {10.1103/PhysRevA.78.022313},
  url = {https://link.aps.org/doi/10.1103/PhysRevA.78.022313}
}

@article{Vacchini_rtn_expt,
  title = {Experimental investigation of the effect of classical noise on quantum non-Markovian dynamics},
  author = {Cialdi, Simone and Benedetti, Claudia and Tamascelli, Dario and Olivares, Stefano and Paris, Matteo G. A. and Vacchini, Bassano},
  journal = {Phys. Rev. A},
  volume = {100},
  issue = {5},
  pages = {052104},
  numpages = {7},
  year = {2019},
  month = {Nov},
  publisher = {American Physical Society},
  doi = {10.1103/PhysRevA.100.052104},
  url = {https://link.aps.org/doi/10.1103/PhysRevA.100.052104}
}

@article{
rtn_expt_2024,
author = {Megan Cowie  and Procopios C. Constantinou  and Neil J. Curson  and Taylor J. Z. Stock  and Peter Grütter },
title = {Spatially resolved random telegraph fluctuations of a single trap at the Si/SiO<sub>2</sub> interface},
journal = {Proceedings of the National Academy of Sciences},
volume = {121},
number = {44},
pages = {e2404456121},
year = {2024},
doi = {10.1073/pnas.2404456121},
URL = {https://www.pnas.org/doi/abs/10.1073/pnas.2404456121}
}

@article{Eli_rtn_2001,
  title = {Time-Dependent Fluctuations in Single Molecule Spectroscopy: A Generalized Wiener-Khintchine Approach},
  author = {Barkai, Eli and Jung, YounJoon and Silbey, Robert},
  journal = {Phys. Rev. Lett.},
  volume = {87},
  issue = {20},
  pages = {207403},
  numpages = {4},
  year = {2001},
  month = {Oct},
  publisher = {American Physical Society},
  doi = {10.1103/PhysRevLett.87.207403},
  url = {https://link.aps.org/doi/10.1103/PhysRevLett.87.207403}
}

@article{Cialdi_2017,
    author = {Cialdi, Simone and Rossi, Matteo A. C. and Benedetti, Claudia and Vacchini, Bassano and Tamascelli, Dario and Olivares, Stefano and Paris, Matteo G. A.},
    title = {All-optical quantum simulator of qubit noisy channels},
    journal = {Applied Physics Letters},
    volume = {110},
    number = {8},
    pages = {081107},
    year = {2017},
    month = {02},
    issn = {0003-6951},
    doi = {10.1063/1.4977023},
    url = {https://doi.org/10.1063/1.4977023}
}

@article{Davidovich_QSL,
  title = {Quantum Speed Limit for Physical Processes},
  author = {Taddei, M. M. and Escher, B. M. and Davidovich, L. and de Matos Filho, R. L.},
  journal = {Phys. Rev. Lett.},
  volume = {110},
  issue = {5},
  pages = {050402},
  numpages = {5},
  year = {2013},
  month = {Jan},
  publisher = {American Physical Society},
  doi = {10.1103/PhysRevLett.110.050402},
  url = {https://link.aps.org/doi/10.1103/PhysRevLett.110.050402}
}

@article{SB_Thomas_NM_heat_engine,
  title = {Thermodynamics of non-Markovian reservoirs and heat engines},
  author = {Thomas, George and Siddharth, Nana and Banerjee, Subhashish and Ghosh, Sibasish},
  journal = {Phys. Rev. E},
  volume = {97},
  issue = {6},
  pages = {062108},
  numpages = {8},
  year = {2018},
  month = {Jun},
  publisher = {American Physical Society},
  doi = {10.1103/PhysRevE.97.062108},
  url = {https://link.aps.org/doi/10.1103/PhysRevE.97.062108}
}

@article{topological_QB,
  title = {Topological Quantum Batteries},
  author = {Lu, Zhi-Guang and Tian, Guoqing and L\"u, Xin-You and Shang, Cheng},
  journal = {Phys. Rev. Lett.},
  volume = {134},
  issue = {18},
  pages = {180401},
  numpages = {8},
  year = {2025},
  month = {May},
  publisher = {American Physical Society},
  doi = {10.1103/PhysRevLett.134.180401},
  url = {https://link.aps.org/doi/10.1103/PhysRevLett.134.180401}
}

@article{Fischer_QB,
    author = {Gyhm, Ju-Yeon and Fischer, Uwe R.},
    title = {Beneficial and detrimental entanglement for quantum battery charging},
    journal = {AVS Quantum Science},
    volume = {6},
    number = {1},
    pages = {012001},
    year = {2024},
    month = {01},
    issn = {2639-0213},
    doi = {10.1116/5.0184903},
    url = {https://doi.org/10.1116/5.0184903}
}

@article{segal_strong_coupling,
  title = {Optimal qubit-mediated quantum heat transfer via noncommuting operators and strong coupling effects},
  author = {Brenes, Marlon and Garwo\l{}a, Jakub and Segal, Dvira},
  journal = {Phys. Rev. B},
  volume = {111},
  issue = {23},
  pages = {235440},
  numpages = {18},
  year = {2025},
  month = {Jun},
  publisher = {American Physical Society},
  doi = {10.1103/m18l-t1hk},
  url = {https://link.aps.org/doi/10.1103/m18l-t1hk}
}

@article{colloquium_QB,
  title = {Colloquium: Quantum batteries},
  author = {Campaioli, Francesco and Gherardini, Stefano and Quach, James Q. and Polini, Marco and Andolina, Gian Marcello},
  journal = {Rev. Mod. Phys.},
  volume = {96},
  issue = {3},
  pages = {031001},
  numpages = {30},
  year = {2024},
  month = {Jul},
  publisher = {American Physical Society},
  doi = {10.1103/RevModPhys.96.031001},
  url = {https://link.aps.org/doi/10.1103/RevModPhys.96.031001}
}

@article{Lidar_central_spin,
  title = {Non-Markovian dynamics of a qubit coupled to an Ising spin bath},
  author = {Krovi, Hari and Oreshkov, Ognyan and Ryazanov, Mikhail and Lidar, Daniel A.},
  journal = {Phys. Rev. A},
  volume = {76},
  issue = {5},
  pages = {052117},
  numpages = {15},
  year = {2007},
  month = {Nov},
  publisher = {American Physical Society},
  doi = {10.1103/PhysRevA.76.052117},
  url = {https://link.aps.org/doi/10.1103/PhysRevA.76.052117}
}

@article{Rashid_2025,
doi = {10.1088/1402-4896/ae06d8},
url = {https://doi.org/10.1088/1402-4896/ae06d8},
year = {2025},
month = {sep},
publisher = {IOP Publishing},
volume = {100},
number = {9},
pages = {095119},
author = {Rashid, Mehboob and Mala, Rayees A and Bashir, Saima and Lone, Muzaffar Qadir},
title = {Non-Markovian amplitude damping in a central spin model with random couplings},
journal = {Physica Scripta}
}

@article{tanimura_exact,
    author = {Tanimura, Yoshitaka},
    title = {Numerically “exact” approach to open quantum dynamics: The hierarchical equations of motion (HEOM)},
    journal = {The Journal of Chemical Physics},
    volume = {153},
    number = {2},
    pages = {020901},
    year = {2020},
    month = {07},
    issn = {0021-9606},
    doi = {10.1063/5.0011599},
    url = {https://doi.org/10.1063/5.0011599}
}

@article{Nori_2024,
  title = {Non-Hermitian pseudomodes for strongly coupled open quantum systems: Unravelings, correlations, and thermodynamics},
  author = {Menczel, Paul and Funo, Ken and Cirio, Mauro and Lambert, Neill and Nori, Franco},
  journal = {Phys. Rev. Res.},
  volume = {6},
  issue = {3},
  pages = {033237},
  numpages = {25},
  year = {2024},
  month = {Sep},
  publisher = {American Physical Society},
  doi = {10.1103/PhysRevResearch.6.033237},
  url = {https://link.aps.org/doi/10.1103/PhysRevResearch.6.033237}
}

@article{Nori_2023,
  title = {Pseudofermion method for the exact description of fermionic environments: From single-molecule electronics to the Kondo resonance},
  author = {Cirio, Mauro and Lambert, Neill and Liang, Pengfei and Kuo, Po-Chen and Chen, Yueh-Nan and Menczel, Paul and Funo, Ken and Nori, Franco},
  journal = {Phys. Rev. Res.},
  volume = {5},
  issue = {3},
  pages = {033011},
  numpages = {16},
  year = {2023},
  month = {Jul},
  publisher = {American Physical Society},
  doi = {10.1103/PhysRevResearch.5.033011},
  url = {https://link.aps.org/doi/10.1103/PhysRevResearch.5.033011}
}
\bibliographystyle{apsrev}
 
\appendix

\newpage
\onecolumngrid
\begin{center}
    \textbf{\large Supplemental Material for ``Finite-Bath Open Quantum Systems: Exact Dynamics''}
\end{center}

\section{The minimal dissipator}
Here, we briefly discuss the minimal dissipation theory~\cite{Hayden_2022}.
Consider a space of the superoperators of the open quantum system containing linear maps $\mathcal{L}: \mathcal{B}(\mathscr{H}) \to \mathcal{B}(\mathscr{H})$. Let the vector space consisting of all superoperators $\mathcal{L}$ be $\mathfrak{qme}(\mathscr{H})$. These superoperators $\mathcal{L}$ satisfy the condition 
$\mathcal{L}[O^\dagger] = \mathcal{L}[O]^\dagger, {\rm Tr}\left\{\mathcal{L}[O]\right\} = 0 ~~\forall O \in \mathcal{B}(\mathscr{H}). 
$
The size of the superoperators in the vector space  $\mathfrak{qme}(\mathscr{H})$ can be quantified using the following norm
$
    ||\mathcal{L}||^2 = \overline{\bra{\psi}\overline{\mathcal{L}[\ket{\phi}\bra{\phi}]^2}\ket{\psi}},
$
where $\ket{\psi}$ and $\ket{\phi}$ are normalized random state vectors. The overline denotes the corresponding average over the Haar measure on the unitary group. This norm is induced by the scalar product
$\langle\mathcal{L}_1,\mathcal{L}_2\rangle = \overline{\bra{\psi}\overline{\mathcal{L}_1[\ket{\phi}\bra{\phi}]\mathcal{L}_2[\ket{\phi}\bra{\phi}]}\ket{\psi}},
$
where $\mathcal{L}_1, \mathcal{L}_2 \in \mathfrak{qme}(\mathscr{H})$. A geometric interpretation of this norm is given in~\cite{breuer_effective_Ham1}. We can define a subspace $\mathfrak{ham}(\mathscr{H})$ of $\mathfrak{qme}(\mathscr{H})$ for which there exists a Hermitian operator $H$ such that $\mathcal{H}[O] = -i[H, O] ~~\forall O \in \mathcal{B}(\mathscr{H})$. Once we have the superoperator $\mathcal{L}_t$, we can define the Hamiltonian part $\mathcal{H}_t$ as the orthogonal projection of $\mathcal{L}_t$ onto the subspace $\mathfrak{ham}(\mathscr{H})$. Thus, we have the following orthogonal projections of the $\mathcal{L}_t$ superoperator: 
$
    \mathcal{H}_t = \Pi(\mathcal{L}_t), \mathcal{D}_t = \mathcal{L}_t - \Pi(\mathcal{L}_t),
$
where $\Pi$ defines the orthogonal projection of $\mathcal{L}_t$ on the $\mathfrak{ham}(\mathscr{H})$ subspace. The above expression results in a unique expression for the Hamiltonian and the dissipator part of the master equation. Moreover, the above decomposition ensures the least norm of the dissipator; therefore, this is termed \textit{the principle of minimal dissipation}. The principle of minimal dissipation ensures that a unique definition of the dissipator of the master equation associated with the heat exchange has the least possible average effect on the physical state, shifting to the highest Hamiltonian (canonical Hamiltonian) contribution. Further, in~\cite{Hayden_2022}, it was demonstrated that a minimal norm of the dissipator exactly corresponds to taking traceless Lindblad jump operators in the master equation. 

The space of the generators of the quantum master equation can be orthogonally decomposed into 
$
    \mathfrak{qme}(\mathscr{H}) = \mathfrak{ham}(\mathscr{H}) \oplus \mathfrak{ham}^{\perp}(\mathscr{H}),
$
such that at any time $t$, the generator $\mathcal{L}_t$ can be written as $\mathcal{L}_t = \mathcal{H}_t + \mathcal{D}_t$, with $\mathcal{H}_t\in \mathfrak{ham}(\mathscr{H})$. To this end, we rewrite the unique decomposition of the generator $\mathcal{L}$ on $\mathfrak{ham}(\mathscr{H})$ as
$
    \mathcal{H}_t = \Pi (\mathcal{L}_t) = \sum_j \mathcal{H}_j \langle\mathcal{H}_j, \mathcal{L}_t\rangle, 
$
where $\left\{\mathcal{H}_j\right\}_{j = 1}^{N^2-1}$ is an orthonormal basis of $\mathfrak{ham}(\mathscr{H})$.  Now, to find the expression for the Hamiltonian $H^{\can}$ generating $\mathcal{H}_t = -i[H^{\can}, \cdot]$, the property  of $\mathfrak{ham}(\mathscr{H})$ being isomorphic to the space of traceless Hermitian operators $\{H \in \mathcal{B}(\mathscr{H}| H^\dagger = H, {\rm Tr} (H) = 0\}$ is exploited. Using this property, the Hamiltonian operator associated with $\mathcal{H}$ is given by
$
    H^{\can} = \sum_j H_j \langle \mathcal{H}_j, \mathcal{L}\rangle.
$
Further, to find the expression for the coefficients $\langle\mathcal{H}_j, \mathcal{L}\rangle$, one can make use of the pseudo-Kraus representation for the generator, which states that any Hermiticity preserving and trace annihilating (HPTA) map can be written as 
\begin{align}
\mathcal{L}\left(\rho\right) = \sum_k\gamma_k E_k\rho E_k^\dagger,
\label{pseudo-Kraus_decomposition}
\end{align}
where $E_k$'s are called the pseudo-Kraus operators and $\gamma_k$'s are real coefficients. Using the above, one can write the Hamiltonian generating the projection of $\mathcal{L}$ on $\mathfrak{ham}(\mathscr{H})$ as 
\begin{align}
    H^{\can} = -\frac{i}{2d}\sum_k\gamma_k\left[{\rm Tr}\left\{E_k\right\}E_k^\dagger - {\rm Tr}\left\{E_k^\dagger\right\}E_k\right],
    \label{canonical_hamiltonian}
\end{align}
where $d$ is the dimension of the reduced system. Correspondingly, the form of the minimal dissipator is given by 
\begin{align}
    \mathcal{D}\left(\rho\right) = \sum_k\gamma_k\left[L_k\rho L_k^\dagger - \frac{1}2\left\{L_k^\dagger L_k, \rho\right\}\right],
    \label{minimal_dissipator}
\end{align}
where the $L_k$ are the Lindblad jump operators given by 
$
    L_k = E_k - \frac{{\rm Tr}\left\{E_k\right\}}{d}\mathbb{I}.
$
Using the canonical Hamiltonian and the minimal dissipator, the master equation takes the following form
\begin{align}
    \frac{d\rho(t)}{dt} = -i[H^{\can}, \rho(t)] + \mathcal{D}\left[\rho(t)\right].
\end{align}

\section{The general dynamical map of a two-level system}
Consider the evolution of the reduced state of a two-level system
\begin{align}\label{eq_dynamics_of_system}
    \rho_S(t) = \varphi_t\left[\rho_S(0)\right],
\end{align}
where $\rho_S(0)$ is given by
$
    \rho_S(0) = \sum_{j, k = 0}^1\rho_{jk}(0)\ket{j}\bra{k},
$
with $\ket{0} \equiv \left(1~~~0\right)^T$ and $\ket{1}\equiv \left(0~~~1\right)^T$, and $\varphi_t$ maps the initial state of the system to the final state, and it is linear. Further, it is a completely positive and trace-preserving (CPTP) map that admits a Kraus decomposition~\cite{KRAUS1971}
\begin{align}\label{eq_app_kraus_ops}
    \varphi_t(X) = \sum_k A_k XA_k^\dagger,
\end{align}
where $A_k$'s are Kraus operators, satisfying $\sum_kA_k^\dagger A_k = \mathbb{I}$. The action of the map on an arbitrary operator $X$, not necessarily a valid density matrix, satisfies the following two fundamental properties. First, it preserves the trace, that is, 
\begin{align}\label{eq_app_trace_preserve}
    \Tr\left[\varphi_t(X)\right] = \Tr\left[\sum_k A_k X A_k^\dagger\right] = \Tr\left[\left(\sum_k A_k^\dagger A_k\right) X\right] = \Tr\left[X\right].
\end{align}
Second, the map preserves Hermiticity, that is, 
\begin{align}\label{eq_app_Herm_preserve}
    \left[\varphi_t(X)\right]^\dagger = \sum_k A_kX^\dagger A_k^\dagger = \varphi_t(X^\dagger).
\end{align}

We now strive to find a general matrix structure of this map for a two-level system. To this effect, we apply the vectorization operation ${\rm vec}\left\{\ket{j}\bra{k}\right\} = \ket{j}\otimes \ket{k} = \kett{jk}$ on Eq.~\eqref{eq_dynamics_of_system} to get
\begin{align}\label{eq_mat_map}
    \kett{\rho_S(t)} = \Phi_t\boldsymbol{\cdot}\sum_{j,k=0}^1\rho_{jk}(0)\kett{jk},
\end{align}
where $\Phi_t$ is the matrix form of the map $\varphi_t$ in the superoperator space and $a\boldsymbol{\cdot}b$ is a multiplication between matrices $a$ and $b$. The elements $\rho_{jk}(0)$ are real coefficients for $j = k$, whereas they are complex for $j\ne k$. Further, they satisfy $\rho_{11}(0) = 1 - \rho_{00}(0)$, and $\rho_{01} = \rho_{10}^*$, where $z^*$ denotes complex conjugate of $z$.  
Now, the right side of Eq.~\eqref{eq_mat_map} can be rewritten as 
\begin{align}\label{eq_app_action_map_RHS}
    \Phi_t\boldsymbol{\cdot}\sum_{j, k =0}^1\rho_{jk}\kett{jk} 
    &= \sum_{j, k =0}^1\rho_{jk}\Phi_t\cdott\kett{jk},
\end{align}
where $\rho_{jk}(0)$ is written as $\rho_{jk}$ for simplicity. We shall examine $\Phi_t\cdott\kett{jk}$ one by one. Let $\Phi_t\cdott\kett{00} = \left(\phi_{11}~~ \phi_{21}~~\phi_{31}~~\phi_{41}\right)^T$, where the elements $\phi_{jk}$ may, in general, be complex and time dependent. Notice that $\kett{00}$, when devectorized, is a Hermitian matrix with trace one, which implies $\phi_{41} = 1 - \phi_{11}$ with $\phi_{11}\in\mathbb{R}$, and $\phi_{31} = \phi_{21}^*$, see Eqs.~\eqref{eq_app_trace_preserve} and~\eqref{eq_app_Herm_preserve}. Thus, $\Phi_t\cdott\kett{00} = \left(\phi_{11}~~ \phi_{21}~~\phi_{21}^*~~1-\phi_{11}\right)^T$. Now, suppose $\Phi_t\cdott\kett{01} = \left(\phi_{12}~~ \phi_{22}~~\phi_{32}~~\phi_{42}\right)^T$, where $\kett{01}$, when devectorized, is non-Hermitian with trace zero that renders $\phi_{42} = -\phi_{12}$. Note that in this case, $\phi_{12}, \phi_{22}$, and $\phi_{32}$ can take complex values. Thus, $\Phi_t\cdott\kett{01} = \left(\phi_{12}~~ \phi_{22}~~\phi_{32}~~-\phi_{12}\right)^T$. Similarly, we can write $\Phi_t\cdott\kett{10} = \left(\phi_{13}~~ \phi_{23}~~\phi_{33}~~-\phi_{13}\right)^T$ and $\Phi_t\cdott\kett{11} = \left(1 - \phi_{44}~~ \phi_{24}~~\phi_{24}^*~~\phi_{44}\right)^T$. Substituting the above in Eq.~\eqref{eq_app_action_map_RHS} and using $\kett{\rho_S(t)} = \left(\rho_{00}(t)~~\rho_{01}(t)~~\rho_{10}(t)~~\rho_{11}(t)\right)^T$, we get
\begin{align}
    \begin{pmatrix}
        \rho_{00}(t) \\ \rho_{01}(t) \\ \rho_{10}(t) \\ \rho_{11}(t)
    \end{pmatrix} = \begin{pmatrix}
        \rho_{00}\phi_{11}+\rho_{01}\phi_{12}+\rho_{10}\phi_{13}+\rho_{11}(1 - \phi_{44})\\
        \rho_{00}\phi_{21}+\rho_{01}\phi_{22}+\rho_{10}\phi_{23}+\rho_{11}\phi_{24}\\
        \rho_{00}\phi_{21}^* +\rho_{01}\phi_{32}+\rho_{10}\phi_{33}+\rho_{11}\phi_{24}^*\\
        \rho_{00}(1 - \phi_{11})-\rho_{01}\phi_{12}-\rho_{10}\phi_{13}+\rho_{11}\phi_{44}\\
    \end{pmatrix}.
\end{align}
The state $\rho_S(t)$ at any time $t$ is Hermitian, that is, $\rho_{01}^*(t) = \rho_{10}(t)$ and $\rho_{00}(t), ~\rho_{11}(t)\in\mathbb{R}$. On applying these properties on the right side of the above equation, we get the following relations: $\phi_{33} = \phi_{22}^*$, $\phi_{32} = \phi_{23}^*$, and $\phi_{13} = \phi_{12}^*$. These relations reduce the number of unknowns in the form of $\phi_{jk}$ in the above equation. Further, the right side of the above equation can be split as $\Phi_t\cdott\kett{\rho_S(0)}$, where the matrix form of the general two-level dynamical map $\Phi_t$ takes the following form
\begin{align}\label{eq_app_dynamical_map_matrix}
    \Phi_t = \begin{pmatrix}
        \phi_{11}&\phi_{12}&\phi_{12}^*&1-\phi_{44}\\
        \phi_{21}&\phi_{22}&\phi_{23}&\phi_{24}\\
        \phi_{21}^*&\phi_{23}^*&\phi_{22}^*&\phi_{24}^*\\
        1-\phi_{11}&-\phi_{12}&-\phi_{12}^*&\phi_{44}
    \end{pmatrix}.
\end{align}

\section{Relationship between the matrix elements of $\mathbb{L}_t$ and $\Phi_t$}
The matrix forms of the Lindbladian superoperator and the general dynamical map of a two-level system are related by $\mathbb{L}_t = \dot\Phi_t\Phi_t^{-1}$. Let the matrix elements of $\Phi_t^{-1}$ be denoted by $\phi^{\rm inv}_{jk}$. In terms of matrix elements of $\Phi_t$, these elements are given by
\begin{align}\label{app_eq_phin_inv_elements}
    \phi^{\rm inv}_{11} &= \frac{1}{\Delta}\left[\phi_{44}\left(\left|\phi_{22}\right|^2 - \left|\phi_{23}\right|^2\right) + 2\Re(\phi_{12}\phi_{24}\phi_{22}^*) + 2\Re(\phi_{12}\phi_{23}\phi_{24}^*)\right], \nonumber \\
    \phi^{\rm inv}_{12} &= \frac{1}{\Delta}\left(\phi_{12}^*\phi_{23}^* - \phi_{12}\phi_{23}^*\right),~~~ \phi^{\rm inv}_{13} = \phi^{\rm inv *}_{12} \nonumber \\
    \phi^{\rm inv}_{14} &= \phi^{\rm inv}_{11} - \frac{1}{\Delta}\left(\left|\phi_{22}\right|^2 - \left|\phi_{23}\right|^2\right), \nonumber \\
    \phi^{\rm inv}_{21} &= \frac{1}{\Delta} \left[(1 - \phi_{11})\left\{\phi_{22}^*\phi_{24} - \phi_{23}\phi_{24}^*\right\} + \phi_{44}\left\{\phi_{21}^*\phi_{23} - \phi_{21}\phi_{22}^*\right\} + \phi_{12}^*\left\{\phi_{24}\phi_{21}^* - \phi_{21}\phi_{24}^*\right\}\right],\nonumber \\
    \phi^{\rm inv}_{22} &= \frac{1}{\Delta}\left[\phi_{22}^*\left(\phi_{11} + \phi_{44} - 1\right) + \phi_{12}^*\left(\phi_{24}^* - \phi_{21}^*\right)\right], \nonumber \\
    \phi^{\rm inv}_{23} &= \frac{1}{\Delta}\left[\phi_{23}\left(1 - \phi_{11} - \phi_{44}\right) + \phi_{12}^*\left(\phi_{21} - \phi_{24}\right)\right], \nonumber \\
    \phi^{\rm inv}_{24} &= \phi^{\rm inv}_{21} + \frac{1}{\Delta} \left[\phi_{22}^*\left(\phi_{21} - \phi_{24}\right) + \phi_{23}\left(\phi_{24}^* - \phi_{21}^*\right)\right], \nonumber \\
    \phi^{\rm inv}_{31} &= \phi^{\rm inv*}_{21},~~~\phi^{\rm inv}_{32} = \phi^{\rm inv*}_{23},~~~\phi^{\rm inv}_{33} = \phi^{\rm inv*}_{22},~~~\phi^{\rm inv}_{34} = \phi^{\rm inv*}_{24}, \nonumber \\
    \phi^{\rm inv}_{41} &= \frac{1}{\Delta}\left[(1 - \phi_{11})\left\{|\phi_{23}|^2 - |\phi_{22}|^2\right\} + 2\Re(\phi_{12}\phi_{21}^*\phi_{23}) - 2\Re(\phi_{12}\phi_{21}\phi_{22}^*)\right], \nonumber \\
    \phi^{\rm inv}_{42} &= -\phi^{\rm inv}_{12}, ~~~\phi^{\rm inv}_{43} = -\phi^{\rm inv}_{13}, \nonumber \\
    \phi^{\rm inv}_{44} &= \phi^{\rm inv}_{41} + \frac{1}{\Delta}\left\{|\phi_{22}|^2 - |\phi_{23}|^2\right\},
\end{align}
where $\Delta$ is the determinant of the matrix $\Phi_t$. 
Using the elements $\phi_{jk}$ of $\Phi_t$ and $\phi^{\rm inv}$ of $\Phi^{-1}_t$, we can write the elements $l_{jk}$ of $\mathbb{L}_t$ as 
\begin{align}
    l_{11} &= \dot \phi_{11}\phi_{11}^\inv + 2\Re\left(\dot \phi_{12}\phi_{21}^\inv\right) - \dot \phi_{44}\phi_{41}^\inv, \nonumber \\
    l_{12} &= \dot\phi_{11} \phi_{12}^\inv + \dot \phi_{12} \phi_{22}^\inv + \dot \phi_{12}^*\phi_{23}^{\inv*} + \dot\phi_{44}\phi_{12}^\inv, \nonumber \\
    l_{13} &= l_{12}^*,\nonumber \\
    l_{14} &= \dot \phi_{11} \phi_{14}^\inv + 2\Re\left(\dot \phi_{12}\phi_{24}^\inv\right) - \dot\phi_{44}\phi_{44}^\inv, \nonumber \\
    l_{21} &= \dot \phi_{21}\phi_{11}^\inv + \dot \phi_{22}\phi_{21}^\inv + \dot \phi_{23}\phi_{21}^{\inv*} + \dot \phi_{24}\phi_{41}^{\inv}, \nonumber \\
    l_{22} &= \dot \phi_{21} \phi_{12}^\inv + \dot \phi_{22}\phi_{22}^\inv + \dot \phi_{23}\phi_{23}^{\inv*} - \dot\phi_{24}\phi_{12}^\inv, \nonumber \\
    l_{23} &= \dot \phi_{21}\phi_{12}^{\inv*} + \dot \phi_{22}\phi_{23}^\inv + \dot \phi_{23} \phi_{22}^{\inv*} - \dot \phi_{24}\dot\phi_{12}^{\inv*},\nonumber \\
    l_{24} &= \dot \phi_{21}\phi_{14}^\inv + \dot\phi_{22}\phi_{24}^\inv + \dot\phi_{23}\phi_{24}^{\inv*} + \dot \phi_{24}\phi_{44}^\inv, \nonumber \\
    l_{31} &= l_{21}^*,~~~l_{32} = l_{23}^*,~~~l_{33} = l_{22}^*,~~~l_{34} = l_{24}^*, \nonumber \\
    l_{41} &= -l_{11},~~~l_{42} = -l_{12},~~~l_{43} = -l_{12}^*,~~~l_{44} = -l_{14}.
\end{align}
The above equation, together with Eq.~\eqref{app_eq_phin_inv_elements}, relates the elements of the map $\Phi_t$ with the elements of the generator $\mathbb{L}_t$ for a general TLS.

\section{Exact solution of the central spin model}
The total Hamiltonian of the central spin model (for $\hbar=1$) is given by~\cite{Chiranjib_2017, Tiwari2025, Devvrat_central_spin1}
\begin{align}\label{SM_eq_CS_H_total}
    H &= H_S + H_B + V =\frac{ \omega_0}{2}\sigma^z + \frac{ \omega}{N} J_z + \lambda (\sigma^xJ_x+\sigma^y J_y),
\end{align}
where $\lambda = \epsilon/\sqrt{N}$, $\omega_0$ is the transition frequency of the central spin-$1/2$ particle, and $\epsilon$ is the interaction strength between the system and the bath, with $\sqrt{N}$ being the scaling factor. $\omega/N$ is the scaled frequency of the bath. The evolution of the composite system $S\text{-}B$ for arbitrary initial states of the system $\rho_S(0)$ and bath $\rho_B(0)$ using the global unitary $U = e^{-iHt}$ is given by 
\begin{align}\label{eq_total_unitary_evolution}
    \rho_{SB}(t) = e^{-iHt}\left\{\rho_S(0) \otimes\rho_B(0)\right\}e^{iHt}.
\end{align}
and the reduced state of the central spin system is given by
\begin{align}
    \rho_S(t)={\rm Tr}_B\left[ e^{-iHt}\left\{\rho_S(0) \otimes \rho_B(0) \right\}e^{iHt}\right],
    \label{SM_Eq_unitary_dynamics}
\end{align}
The thermal Gibbs state is taken as the initial state of the bath, which is given by
\begin{align}\label{rhoB0}
    \rho_B(0)&= \frac{e^{-\beta H_B}}{Z}  =\frac{1}{Z}\sum_{n=0}^{N}e^{-\frac{\beta\omega}{2}\left(1-\frac{2n}{N}\right)}\ket{n}\bra{n},
\end{align}
where $\beta = 1/T$ (for $k_B =1$), $\ket{n}$ is the standard computational basis, and $Z = \sum_{n=0}^{N}e^{-\frac{\beta\omega}{2}\left(1-\frac{2n}{N}\right)}$ is the partition function. 
We now derive the spectral decomposition of the total Hamiltonian $H$, which is given by
\begin{align}
    H &= \sum_{n=1}^N \lambda_{\pm}(n) \ket{\psi_{\pm}(n)}\bra{\psi_{\pm}(n)} + \left(\frac{\omega + \omega_0}{2}\right)\ket{0}_S\ket{0}_B\bra{0}_S\bra{0}_B - \left(\frac{\omega + \omega_0}{2}\right)\ket{1}_S\ket{N}_B\bra{1}_S\bra{N}_B,
\end{align}
where 
\begin{align}
    \ket{\psi_\pm(n)} = \frac{\chi_\pm(n) \ket{0}_S \ket{n}_B + \ket{1}_S \ket{n-1}_B}{\sqrt{1 + \chi_\pm (n)^2}},
\end{align}
and 
\begin{align}
    \lambda_\pm (n) = \frac{\left\{N - (2n - 1)\right\}\omega \pm \sqrt{b^2 + 4a_n^2}}{2N}. 
\end{align}
$\ket{i}$ is the standard computational basis in the above equations. The factor $\chi_\pm(n)$ is given by
\begin{align}
    \chi_{\pm}(n)&=\frac{-b\pm \sqrt{b^2+4a_n^2}}{2a_n},
\end{align}
where $b=\omega-N\omega_0$, and 
\begin{align}
    a_n&=\epsilon\sqrt{N}\sqrt{\frac{N}{2}\biggl(\frac{N}{2}+1\biggr)-\biggl(-\frac{N}{2}+n-1\biggr)\biggl(-\frac{N}{2}+n\biggr)}.
\end{align}
Here, $N$ is the number of bath spins, and $n$ is the eigenvalue (eigenvector) index in the range $1\leq n \leq N$. We use these eigenvalues and eigenvectors in Eq. (\ref{SM_Eq_unitary_dynamics}) to obtain the exact dynamics of the central spin system. For an arbitrary initial state $\rho_S(0) = \begin{pmatrix}
    \rho_{00}(0) && \rho_{01}(0)\\\rho_{10}(0) && \rho_{11}(0)
\end{pmatrix}$ of the system, the density matrix governing the evolution of the system is given by
\begin{align}\label{rhos(t)}
    \rho_S(t)=\begin{pmatrix}
        \alpha_t\rho_{00}(0)+\eta_t\rho_{11}(0) && \delta_t\rho_{01}(0) \\
        \delta^*_t\rho_{10}(0) &&  1-\alpha_t\rho_{00}(0)-\eta_t\rho_{11}(0)    
    \end{pmatrix},
\end{align}
where 
\begin{align}
    \alpha_t &= \frac{e^{-\frac{\beta \omega}{2}}}{Z}\left[1 + \sum_{n = 1}^N e^{\frac{n\beta \omega}{N}}\left\{A_n^2 + B_n^2 + 2A_nB_n\cos(\theta_n t)\right\}\right],
\end{align}
with $A_n = \frac{\chi_+(n)^2}{1+\chi_+(n)^2}$, $B_n = \frac{\chi_-(n)^2}{1+\chi_-(n)^2}$, and $\theta_n = \frac{1}{N}\sqrt{b^2 + 4a_n^2}$. Further, in the above equation,
\begin{align}\label{eq_etat}
\eta_t&=\frac{e^{\frac{-\beta\omega}{2}}}{Z}\sum^N_{n=1}e^{\frac{(n-1)\beta\omega}{N}} \left[C_n^2 + D_n^2 + 2C_nD_n\cos(\theta_n t)\right],
\end{align}
where $C_n = \frac{\chi_+(n)}{1+\chi_+(n)^2}$ and $D_n = \frac{\chi_-(n)}{1+\chi_-(n)^2}$. The off-diagonal term $\delta^*_t$ is given by
\begin{align}\label{SM_eq_delta_t}
    \delta^*_t&= \frac{1}{Z}\left[\Xi(t)f_1(t) + \Xi^{-1}(t)h_N(t)\right] + \frac{e^{-\beta\omega/2}}{Z}\sum_{n=2}^N e^{(n-1)\beta\omega/N} f_n(t) g_{n-1}(t),
\end{align}
where $f_n(t) = \frac{e^{-i\lambda_+(n)t}}{1+\chi_+(n)^2}+\frac{e^{-i\lambda_-(n)t}}{1+\chi_-(n)^2}$, $g_n(t) = e^{i\lambda_+(n)t}\frac{\chi_+(n)^2}{1+\chi_+(n)^2}+e^{i\lambda_-(n)t}\frac{\chi_-(n)^2}{1+\chi_-(n)^2}$, $h_N = e^{i\lambda_+(N)t}\frac{\chi_+(N)^2}{1+\chi_+(1)^2}+e^{i\lambda_-(N)t}\frac{\chi_-(N)^2}{1+\chi_-(1)^2}$, and $\Xi(t) = e^{i(\omega + \omega_0)t/2}~e^{-\beta \omega/2}$. These expressions have been used in this work and are very convenient for finding the state of the system for large bath sizes. 

\subsection{Kraus operators for the exact dynamics of the central spin model}
A dynamical map of an open quantum system model can be written in an operator-sum representation~\cite{KRAUS1971, BreurPoqs, banerjee2018open} 
\begin{align}
    \rho_S(t) = \sum_jK_j(t)\rho_S(0)K_j(t)^\dagger,
\end{align}
where the operators $K_j(t)$ are called the Kraus operators. They satisfy the property $\sum_jK_j(t)^\dagger{}K_j(t) = \mathbb{I}_S$. Here, we derive the Kraus operators for the central spin model's dynamics.
Consider the exact dynamical map of the central spin model 
\begin{align}
    \Phi(t) = \begin{pmatrix}
        \alpha_t&0&0&\eta_t\\
        0&\delta_t&0&0\\
        0&0&\delta^*_t&0\\
        1 - \alpha_t&0&0&1 - \eta_t
    \end{pmatrix},
\end{align}
where $\alpha_t, \eta_t$, and $\delta_t$ are defined above. The Kraus operators can be found using the eigenvalues and eigenvectors of the corresponding Choi matrix $C = \left(\Phi(t)\otimes\mathbb{I}_S\right)\ket{\Psi}\bra{\Psi}$, where $\ket{\Psi} = \sum_{i = 1}^d\ket{ii}$ is the maximally entangled state in the Hilbert space of dimension $d^2$. The Choi matrix is given by
\begin{align}
    \mathcal{C}_{\rm map}(t) = \begin{pmatrix}
        \alpha_t&0&0&\delta_t\\
        0&\eta_t&0&0\\
        0&0&1-\alpha_t&0\\
        \delta^*_t&0&0&1-\alpha_t
    \end{pmatrix}.
\end{align}
The eigenvalues $\lambda_i$ and eigenvectors $v_i$ (for $i = 1, 2, 3, 4$) of the above Choi matrix are $\lambda_1 = \eta_t, v_1 = (0~~1~~0~~0)^T$; $\lambda_2 = 1 - \alpha_t, v_2 = (0~~0~~1~~0)^T$; $\lambda_3= \frac{1}{2}\left[\alpha_t + 1-\eta_t + \sqrt{(\alpha_t + \eta_t-1)^2 + 4|\delta_t|^2}\right], v_3= \frac{1}{\sqrt{1 + |x_+|^2}}(x_+~~0~~0~~1)^T$; and $\lambda_4= \frac{1}{2}\left[\alpha_t + 1-\eta_t - \sqrt{(\alpha_t + \eta_t-1)^2 + 4|\delta_t|^2}\right], v_4= \frac{1}{\sqrt{1 + |x_-|^2}}(x_-~~0~~0~~1)^T$, where $x_\pm = \frac{1}{2\delta^*_t}\left[\alpha_t + \eta_t - 1\pm \sqrt{(\alpha_t + \eta_t -1)^2 + 4|\delta_t|^2}\right]$. Using these eigenvalues and eigenvectors, the Kraus operators are given by
\begin{align}
    K_1(t) &= \sqrt{\eta_t}\sigma^+, && K_2(t) = \sqrt{1 - \alpha_t}\sigma^-,
    && K_3(t) = \frac{\sqrt{\lambda_3}}{\sqrt{1 + |x_+|^2}}\begin{pmatrix}
        x_+&0\\0&1
    \end{pmatrix}, && \text{and} && K_4(t) = \frac{\sqrt{\lambda_4}}{\sqrt{1 + |x_-|^2}}\begin{pmatrix}
        x_-&0\\0&1
    \end{pmatrix}. 
\end{align}
These Kraus operators characterize the quantum channel corresponding to the exact dynamics of the central spin model. 

\subsection{Comparison between the perturbative and exact canonical Hamiltonian of the system}
Here, we discuss and compare the canonical Hamiltonian obtained using the perturbative technique discussed in~\cite{Hayden_2022, colla2025_perturbation} with the exact canonical Hamiltonian for the dissipative central spin model. We briefly review the perturbation technique below. Consider the following time-independent Hamiltonian for two finite-dimensional systems $S$ and $B$
\begin{align}\label{supp_eq_total_Ham}
    H = H_S + H_B + H_{SB},
\end{align}
where $H_{SB}$ can be written as $\lambda V$; $V$ being an operator that acts on both $S$ and $B$ with interaction strength $\lambda$. The evolution of the system $S$ for a composite initial state $\rho_S(0)\otimes \rho_B(0)$ is given by
\begin{align}
    \rho_S(t) = \Tr_B\left[U \rho_S(0)\otimes \rho_B(0) U^\dagger{}\right],
\end{align}
where $U = e^{-i(H_S + H_B + \lambda V)t}$. The evolution of the system can be effectively denoted by a dynamical map $\Phi(t)$, such that 
\begin{align}
    \rho_S(t) = \Phi(t)\left[\rho_S(0)\right].
\end{align}
Further, the master equation generating the time evolution of the system $S$ is given by 
\begin{align}
    \frac{\der \rho_S(t)}{\der t} = \mathcal{L}_t\rho_S(t).
\end{align}
\begin{figure}
    \centering
    \includegraphics[width=1\linewidth]{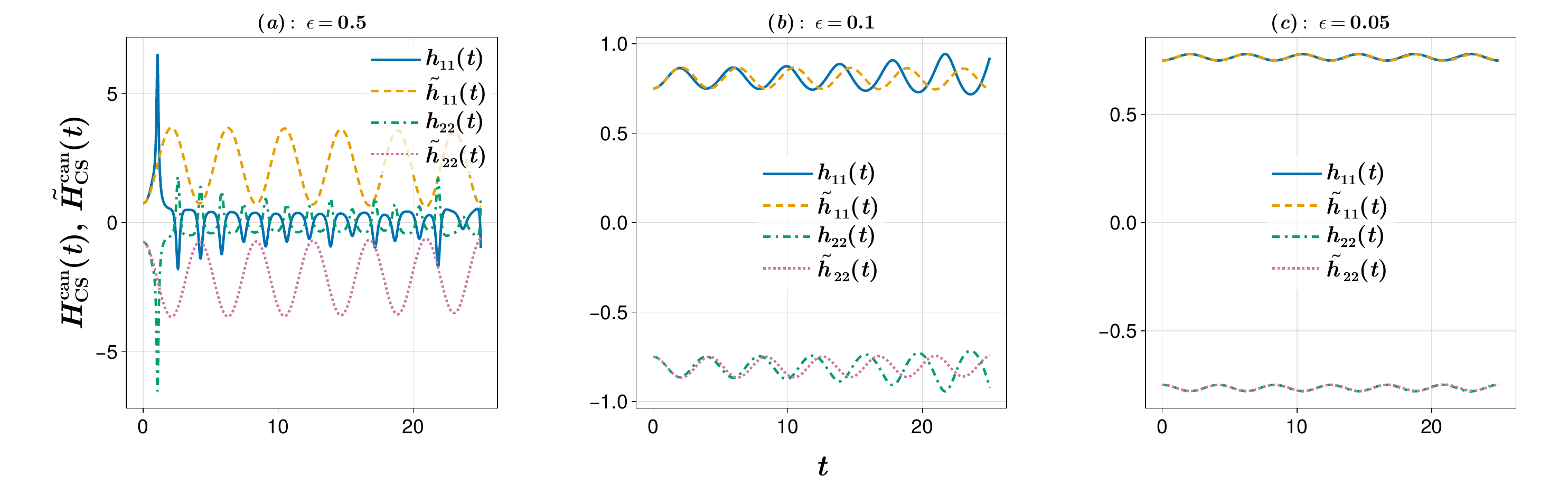}
    \caption{Variation of the matrix elements $h_{jk}(t)$ of the canonical Hamiltonian $H^{\can}_\cs(t)$ obtained using exact dynamics, and the elements $\tilde h_{jk}$ of the canonical Hamiltonian $\tilde H^{\can}_\cs(t)$ obtained using perturbative method. The interaction strength $\lambda = \epsilon/\sqrt{N}$ is varied across the subplots. In (a), $\epsilon = 0.5$, in (b) $\epsilon = 0.1$, and in (c) $\epsilon = 0.05$. The parameters are taken to be: $\omega = 1.0, \omega_0 = 1.5, N = 50$, and $\beta = 0.5$.}
    \label{supp_fig_comparison_exact_and_perturb_canHam}
\end{figure}
For weak $\lambda$, the dynamical map of the system admits a convergent Taylor series of superoperators
\begin{align}
    \Phi(t) = \Phi(t)^{(0)} + \lambda \Phi(t)^{(1)} + \lambda^2 \Phi(t)^{(2)} + \dots~.
\end{align}
A similar expansion can be done for $\mathcal{L}_t$, such that
\begin{align}
    \mathcal{L}_t = \mathcal{L}_t^{(0)} + \lambda \mathcal{L}_t^{(1)} + \lambda^2\mathcal{L}_t^{(2)} + \dots~~.
\end{align}
The terms $\mathcal{L}_t^{(k)}$, in general, are given by~\cite{Hayden_2022} 
\begin{align}
    \mathcal{L}_t^{(k)}[\rho_S(t)] = \frac{\der \Phi(t)^{(k)}}{\der t}\left(e^{iH_St}\rho_S(t)e^{-iH_St}\right) - \sum_{m = 0 }^{k - 1}\left(\mathcal{L}_t^{(m)}\circ\Phi(t)^{(k-m)}\right) \left(e^{iH_St}\rho_S(t)e^{-iH_St}\right).
\end{align}
Truncating the series up to the second order and for the Hamiltonian in Eq.~\eqref{supp_eq_total_Ham}, we can write
\begin{align}
    \mathcal{L}_t^{(0)}[\rho_S(t)] &= -i[H_S, \rho_S(t)], && \mathcal{L}^{(1)}_t[\rho_S(t)] = -i[\Tr_B\left\{V\rho_B(0)\right\}, \rho_S(t)],&&\text{and} &&
    \mathcal{L}_t^{(2)}[\rho_S(t)] = \Tr_B\left[V - \hat V\otimes\mathbb{I}_B, \left[\rho_S(t)\otimes\rho_B(0), \tilde V(t)\right]\right],
\end{align}
where $\hat V = \Tr_B[V\rho_B(0)]$, and $\tilde V(t) = e^{-iH_S t}\left(\int_0^tdsV_{\text{int}}(s)\right)e^{iH_S t}$ with $V_{\text{int}}(t) = e^{i(H_S + H_B) t} V e^{-i (H_S + H_B) t }$. Now, from $\mathcal{L}_t^{(2)}$ using $\tilde V_B(t) = \Tr_S[\tilde V(t)]$ and $V_B = \Tr_S[V]$, the second-order correction to the Hamiltonian of the system can be obtained as
\begin{align}
    H_S^{(2)}(t) = \frac{1}{2id}\Tr_B\left(\tilde V(t)[V_B, \rho_B(0)]\right) + \frac{1}{2id}\Tr_B\left(V[\tilde V_B(t), \rho_B(0)]\right) + \frac{1}{2i}\Tr_B\left([V, \tilde V(t)]\rho_B(0)\right) + \frac{1}{2i}\left[\Tr_B\left\{\tilde V(t)\rho_B(0)\right\}, \hat V\right] + \alpha \mathbb{I}_S,
\end{align}
where $d$ is the dimension of the system. Further, the term $\alpha\mathbb{I}_S$ does not affect the dynamics. The series expansion in terms of $\lambda$ of the canonical Hamiltonian thus becomes 
\begin{align}
    \tilde H^{\can}(t) = H^{(0)}_S + \lambda H^{(1)}_S + \lambda^2H^{(2)}_S(t) + \dots~~,
\end{align}
where $H^{(0)}_S = H_S$, and $H_S^{(1)} = \Tr_B[V\rho_B(0)]$. To get the perturbative canonical Hamiltonian for the central spin model, we use Eq.~\eqref{SM_eq_CS_H_total} in the above equation, and take $\rho_B(0) = e^{-\beta H_B}/\Tr[e^{-\beta H_B}]$, Eq.~\eqref{rhoB0}. It can be shown that for the central spin model $H^{(1)}_S = 0$, and 
\begin{align}
    H^{(2)}_S(t) = c(t)S_1\left[e^{\beta\omega/N}\ket{0}\bra{0} - \ket{1}\bra{1}\right], 
\end{align}
where 
\begin{align}
c(t) = N\left[\cos\left(\omega t/N\right) - \cos\left(\omega_0 t\right)\right]/(N\omega_0 - \omega) && \text{and} && 
S_1 = -e^{\beta\omega/(2N)}\frac{\left[(N+2)\sinh(\beta\omega/2) - N\sinh\left(\beta\omega(N+2)/2N\right)\right]}{\sinh(\beta\omega(N+1)/2N)(e^{\beta\omega/N} - 1)^2}.
\end{align}
From this, we get the perturbative canonical Hamiltonian up to the second order of $\lambda$ for the central spin model as 
\begin{align}
\tilde H^{\can}_\cs(t) = H_S + \lambda^2 H^{(2)}_S(t) = \frac{\omega_0}{2}\sigma^z + \lambda^2 c(t)S_1\left[e^{\beta\omega/N}\ket{0}\bra{0} - \ket{1}\bra{1}\right].
\end{align}
A comparison between the exact canonical Hamiltonian obtained in the main text $H^\can_\cs(t) = \frac{-\Im(\Theta_t)}{2}\sigma_z$ (where $\Theta_t = \frac{\dot\delta_t}{\delta_t}$, Eq.~\eqref{SM_eq_delta_t}, and $\Im(z)$ denotes imaginary part of $z$) and the perturbative canonical Hamiltonian $\tilde H_\cs^{\can}(t)$ is depicted in Fig.~\ref{supp_fig_comparison_exact_and_perturb_canHam}.  It can be observed that, at higher coupling strengths, the exact $H^\can_\cs(t)$ diverges significantly from the perturbative one. This highlights the inadequacy of perturbative methods in this regime, particularly for the strong coupling scenario. Hence, the exact approach becomes necessary. However, as the coupling strength is decreased, the perturbative and exact canonical Hamiltonians converge towards similar values, even overlaying exactly at the low coupling value of $\epsilon = 0.05$.

\subsection{Impact of number of bath spins and interaction strength on charging power and heat current}
Here, we observe the impact of the number of bath spins $N$ and the system bath coupling strength $\epsilon$ on the charging power, heat current, and passive heat current (see Sec. IV of the main manuscript), using the exact master equation developed in the main text for dissipative dynamics of the central spin model, see Eq.~\eqref{eq_CS_master_eq} of the main manuscript. 
\begin{figure}
    \centering
    \includegraphics[width=0.75\linewidth]{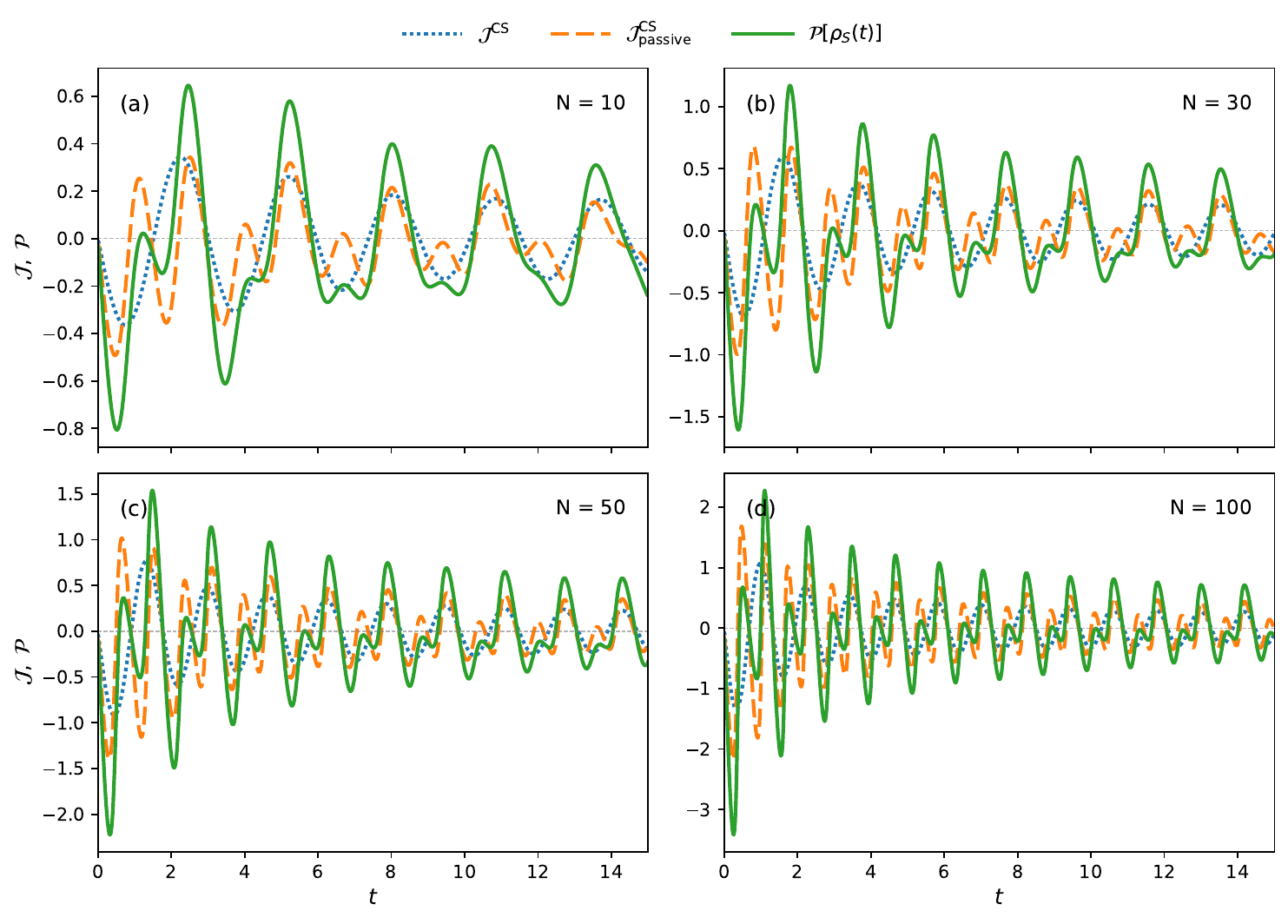}
    \caption{Variation of the heat current, $\mathcal{J}^\cs$, passive heat current $\mathcal{J}^\cs_{\rm passive}$, and charging power $\mathcal{P}[\rho_S(t)]$ for the dissipative central spin model, Eq.~\eqref{eq_CS_Ham}, for different values of bath spins (a): $N = 10$, (b): $ N = 30$, (c): $N = 50$, (d): $N = 100$. Other parameters are: $\omega=1.0, \omega_0=1.5, \epsilon=0.5, \beta=0.5$, and the initial state is taken to be: $\ket{\psi_S(0)} = \frac{\sqrt{3}}{2}\ket{0} + \frac{1}{2}\ket{1}$.}
    \label{supp_fig_plot_vary_N}
\end{figure}
\begin{figure}
    \centering
    \includegraphics[width=0.75\linewidth]{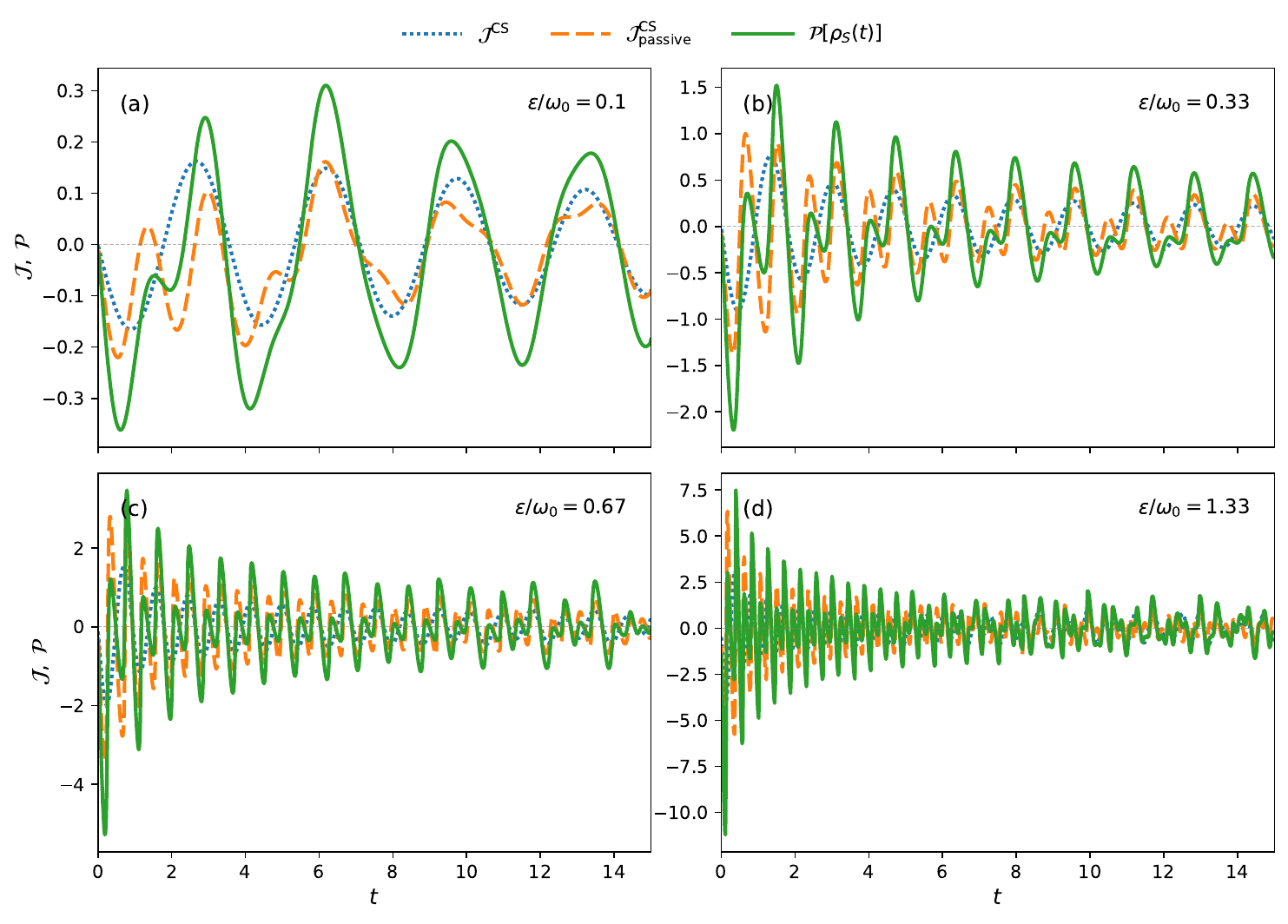}
    \caption{Variation of the heat current, $\mathcal{J}^\cs$, passive heat current $\mathcal{J}^\cs_{\rm passive}$, and charging power $\mathcal{P}[\rho_S(t)]$ for the dissipative central spin model, Eq.~\eqref{eq_CS_Ham}, for different values of coupling strength (a): $\epsilon/\omega_0 = 0.1$, (b): $\epsilon/\omega_0 = 0.33$, (c): $\epsilon/\omega_0 = 0.67$, (d): $\epsilon/\omega_0 = 1.33$. Other parameters are: $\omega=1.0, \omega_0=1.5, N = 50, \beta=0.5$, and the initial state is taken to be: $\ket{\psi_S(0)} = \frac{\sqrt{3}}{2}\ket{0} + \frac{1}{2}\ket{1}$.}
    \label{supp_fig_plot_vary_eps_ratio}
\end{figure}

In Fig.~\ref{supp_fig_plot_vary_N}, we observe the variation of the heat current, $\mathcal{J}^\cs$, passive heat current $\mathcal{J}^\cs_{\rm passive}$, and charging power $\mathcal{P}[\rho_S(t)]$ for different values of bath spins. For the smaller bath size $N=10$, the heat current, passive heat current, and charging power exhibit low-amplitude oscillations. As $N$ increases from 30 to 50, the charging power amplitude grows from around 1.5 to around 2.0, with the oscillation frequency increasing as the denser bath spectral density drives richer non-Markovian recurrences. At $N=100$, the charging power amplitude reaches around 3.0 with approximately 40 sign reversals over $t\in[0,15]$, as more incommensurate bath frequencies contribute to the dynamics. The persistent oscillatory structure across all values of $N$, without any visible decay, confirms that the finite-bath recurrence time remains within the observation window, distinguishing the finite-bath dynamics from the infinite-bath thermodynamic limit.

The variation of the heat current, $\mathcal{J}^\cs$, passive heat current $\mathcal{J}^\cs_{\rm passive}$, and charging power $\mathcal{P}[\rho_S(t)]$ for different values of coupling strength $\epsilon$ is plotted in Fig.~\ref{supp_fig_plot_vary_eps_ratio}. It can be observed that in the weaker coupling regime $\epsilon/\omega_0 = 0.1$, the heat current, passive heat current, and charging power remain small in magnitude and oscillate slowly, consistent with minimal energy exchange when the system-bath interaction is weaker than the bare system energy scale. As the coupling is increased to $\epsilon/\omega_0 = 0.33$ and 0.67, the charging power amplitude grows from around 2.0 to around 5, and the oscillation frequency increases substantially, signaling the onset and deepening of non-Markovian memory effects as coherent energy back-flow from the bath becomes significant. At the strongest coupling $\epsilon/\omega_0 = 1.33$, the charging power amplitude exceeds 10 with many more sign reversals, indicating highly oscillatory dynamics in which energy flows back and forth between the system and bath on timescales comparable to $1/\omega_0$. This strongly coupled regime is inaccessible to perturbative or Markovian treatments, underscoring the necessity of the exact approach developed in this work.

\section{Random Telegraph Noise Master Equation}
\subsection{Derivation of random telegraph noise (RTN) dephasing factor}
In the main text, the RTN master equation is derived using the following variation of the central spin system
\begin{align}
    H &= H_S + H_B + V =\frac{ \omega_0}{2}\sigma^z + \frac{ \omega}{N} J_z + \frac{\epsilon(t)}{\sqrt{N}} \sigma^zJ_z,
\end{align}
which renders a pure dephasing evolution of the central spin system given by 
\begin{align}\label{SM_eq_RTN_density_matrix}
    \rho_S^{01}(t) = \rho_S^{01}(0)e^{-i\omega_0 t}\Lambda(t),
\end{align}
where
\begin{align}\label{SM_eq_Lambdat_1}
    \Lambda(t) = \Tr\left[e^{-\frac{2i}{\sqrt{N}}J_z\int_0^t \epsilon(s)ds}\rho_B(0)\right].
\end{align}
Upon taking the ground state, $\rho_B(0) = \ket{m_0}\bra{m_0}$, where $\ket{m_0} = (0~~0~~\dots 0~~1)_{1\times (N+1)}^T$, as the initial state of the bath and assuming that the time-dependent $\epsilon(t)$ originates from a stochastic process, the dephasing factor $\Lambda(t)$ is given by 
\begin{align}
    \Lambda(t) = \left\langle e^{i\sqrt{N}\int_0^t\epsilon(s) ds}\right\rangle_\epsilon,
\end{align}
where $\langle\cdot\rangle_\epsilon$ denotes an average over the stochastic process characterized by $\epsilon$. For a random telegraph process,
we assume $\epsilon(t)$ switches between $\pm b = \pm\epsilon\sqrt{N}$ randomly with rates $\gamma_{\pm}$ and $\gamma_{\mp}$, where $\gamma_{\pm}(\gamma_{\mp})$ denotes a switch from $+b(-b)$ to $-b(+b)$. A symmetric random telegraph process has rates $\gamma_{\pm} = \gamma_{\mp} = \gamma$. The average defined in Eq.~\eqref{SM_eq_Lambdat_1} can be decomposed into the following conditional averages
\begin{align}\label{SM_eq_cond_Lambda1}
    \Lambda_+(t) = \left\langle e^{i\sqrt{N}\int_0^t\epsilon(s)ds}\right\rangle_{\epsilon|\epsilon(t) = +\epsilon}, && \text{and}&&\Lambda_-(t) = \left\langle e^{i\sqrt{N}\int_0^t\epsilon(s)ds}\right\rangle_{\epsilon|\epsilon(t) = -\epsilon}.
\end{align}
During a short interval \([t, t+dt]\), the RTN process either stays in the same state with probability \(1-\gamma dt\), or flips with probability \(\gamma dt\). The waiting time between successive flips is exponentially distributed with parameter $\gamma$, which comes from the Poisson distribution underlying the switching. The integration $\sqrt{N}\int_0^{t+dt}\epsilon(s)ds$ can be broken into $\sqrt{N}\int_0^t\epsilon(s) ds + \sqrt{N}\int_{t}^{t + dt}\epsilon(s)ds$ and the factor $e^{i\sqrt{N}\int_0^{t + dt}\epsilon(s) ds}$ becomes $e^{i\sqrt{N}\int_0^t\epsilon(s) ds}e^{i\sqrt{N}\int_t^{t+dt}\epsilon(s) ds}$. In time $t$ to $t+dt$, $\epsilon(s)$ can take either $+\epsilon$ or $-\epsilon$ value and thus we can write $\sqrt{N}\int_{t}^{t + dt}\epsilon(s) ds = +\sqrt{N}\epsilon dt = +bdt$ or $-\sqrt{N}\epsilon dt = -bdt$. Using this, the conditional averages $\Lambda_+$ and $\Lambda_-$, Eq.~\eqref{SM_eq_cond_Lambda1}, are given at time $t+dt$ by
\begin{align}
\Lambda_{+}(t+dt) = (1-\gamma dt) e^{+ib dt} \Lambda_{+}(t) + (\gamma dt) e^{-ib dt}\Lambda_{-}(t), && \text{and} &&
\Lambda_{-}(t+dt) = (1-\gamma dt) e^{-ib dt} \Lambda_{-}(t) + (\gamma dt) e^{+ib dt}\Lambda_{+}(t).
\end{align}
Expanding $\Lambda_+(t + dt)$ in the first order of $dt$, we can write
\begin{align}
    \Lambda_+(t + dt) &= (1 - \gamma dt)\left[1 + ibdt + \mathcal{O}(dt^2)\right]\Lambda_+(t) + \gamma dt \left[1 - ibdt + \mathcal{O}(dt^2)\right]\Lambda_-(t), \\ \nonumber
    &= \Lambda_+(t) + (ib - \gamma)dt\Lambda_+(t) + \gamma dt\Lambda_-(t) + \mathcal{O}(dt^2). 
\end{align}
Subtracting \(\Lambda_{+}(t)\) on both sides of the above equation, dividing by \(dt\), and taking the limit \(dt \to 0\), we obtain
\begin{align}
\dot{\Lambda}_{+}(t) = +ib \Lambda_{+}(t) - \gamma \Lambda_{+}(t) + \gamma \Lambda_{-}(t).
\end{align}
Similarly, we can write 
\begin{align}
\dot{\Lambda}_{-}(t) = -ib \Lambda_{-}(t) - \gamma \Lambda_{-}(t) + \gamma \Lambda_{+}(t).
\end{align}
Defining the averaged decoherence factor
\(
\Lambda(t) = \frac{1}{2}\left[ \Lambda_{+}(t) + \Lambda_{-}(t) \right],
\)
one can eliminate variables to show that \(\Lambda(t)\) satisfies
\begin{align}\label{SM_eq_ode_for_Lambdat}
\ddot{\Lambda}(t) + 2\gamma \dot{\Lambda}(t) + b^2 \Lambda(t) = 0.
\end{align}
The general solution of the above ordinary differential equation is given by $\Lambda(t) = C_1e^{m_1}t + C_2e^{m_2t}$, where $m_{1, 2}$ are the roots of the characteristic polynomial: $m^2 + 2\gamma m + b^2 = 0$. In terms of $b$ and $\gamma$, $m_1$ and $m_2$ are given by $m_1 = -\gamma + \sqrt{\gamma^2 - b^2}$ and $m_2 = -\gamma - \sqrt{\gamma^2 - b^2}$. Defining $\mu = \sqrt{\left(\frac{b}{\gamma}\right)^2 - 1}$, and using the initial conditions $\Lambda(0) =1$ and $\dot\Lambda(0) = 0$, the solution of Eq.~\eqref{SM_eq_ode_for_Lambdat} is given by
\begin{align}\label{SM_eq_RTN_Lambda_final_form}
    \Lambda(t) = \begin{cases}
        e^{-\gamma t}\left[\cos\left(\mu\gamma t\right) + \frac{1}{\mu}\sin(\mu\gamma t)\right], &\text{if}~~\left(\frac{b}{\gamma}\right)^2>1, \\
        e^{-\gamma t} (1 + \gamma t), &\text{if}~~b=\gamma,~~\text{and}\\
        e^{-\gamma t}\left[\cosh\left(\mu\gamma t\right) + \frac{1}{\mu}\sinh(\mu\gamma t)\right], &\text{if}~~\left(\frac{b}{\gamma}\right)^2<1.
    \end{cases}
\end{align}
\subsection{Kraus operators  characterizing the RTN model}
The dynamical map for the evolution described in Eq.~\eqref{SM_eq_RTN_density_matrix} is given by 
\begin{align}
    \Phi_t^\rtn = \begin{pmatrix}
        1&0&0&0\\
        0&e^{-i\omega_0 t}\Lambda(t)&0&0\\
        0&0&e^{i\omega_0 t}\Lambda(t)&0\\
        0&0&0&1
    \end{pmatrix}.
\end{align}
The eigenvalues and the corresponding eigenvectors of this Choi Matrix are $1 + \Lambda(t), \frac{1}{\sqrt{2}}\left(e^{-i\omega_0 t/2}~~0~~0~~e^{i\omega_0t/2}\right)^T$ and $1 - \Lambda(t), \frac{1}{2}\left(e^{-i\omega_0 t/2}~~0~~0~~-e^{i\omega_0 t/2}\right)^T$. From these eigenvalues and eigenvectors, the Kraus operators for the evolution of the density matrix in the RTN model are given by 
\begin{align}
    K_1(t)^\rtn = \sqrt{\frac{1 + \Lambda(t)}{2}}U_S(t)\mathbb{I}_S,&&\text{and}&& K_2(t)^\rtn = \sqrt{\frac{1 - \Lambda(t)}{2}}U_S(t)\sigma^z,
    \label{eq_RTN_Kraus_ops}
\end{align}
where $U_S(t) = \exp(-i\omega_0\sigma^zt/2)$. It can be verified that the above Kraus operators follow $\sum_jK_j(t)^{\rtn\dagger} K_j(t)^\rtn = \mathbb{I}_S$ and the system's density matrix at any time $t$ is given by $\rho_S(t) = \sum_jK_j(t)^\rtn\rho_S(0)K_j(t)^\rtn$. 
In the interaction picture, the above Kraus operators take the following form
\begin{align}
    \tilde K_1(t)^\rtn = U_S(t)^\dagger K_1(t)^\rtn = \sqrt{\frac{1 + \Lambda(t)}{2}}\mathbb{I}_S, &&\text{and} && \tilde K_2(t)^\rtn = U_S(t)^\dagger K_2(t)^\rtn = \sqrt{\frac{1 - \Lambda(t)}{2}}\sigma^z.
\end{align}
These exactly match those obtained in~\cite{rtn_model_1} for the RTN model, where the system under consideration was a two-level atom interacting with a fluctuating magnetic field.
Hence, we are able to provide a microscopic derivation of this model, originally constructed using a stochastic Hamiltonian and derived in a semi-classical fashion. This model enjoys immense popularity among the noise channels employed to study non-Markovian dynamics~\cite{rtn_model_1, SO_rice, Bergli_2006_2, RTN1_Bergli_2009, Joynt_2008, RTN3_Cai2020, RTN_dissipative}, and has been experimentally realized, for example, in~\cite{Eli_rtn_2001, Vacchini_rtn_expt, rtn_expt_2024, Paladino_2002, Cialdi_2017}.

\subsection{On incompatibility of the bosonic bath in producing RTN evolution}
Consider the following Hamiltonian for the evolution of a two-level system
\begin{align}\label{SM_eq_total_bosonic_Hamiltonian}
    H = H_S + H_B + H_{SB} = \frac{\omega_0}{2}\sigma^z + \sum_k\omega_ka_k^\dagger a_k + \sigma^z\sum_kg_k(a^\dagger_k + a_k),
\end{align}
where $\omega_0$ is the transition frequency for the two-level atom, and $\omega_k$'s represent the modes of the bosonic bath. To compare with the RTN model's dynamics, we consider an initial vacuum state of the bath $\ket{0_B}$, and an initial product state of the composite system and bath $\rho_{SB}(0) = \rho_S(0)\otimes|0_B\rangle\langle 0_B|$. The unitary operator corresponding to the total Hamiltonian in Eq.~\eqref{SM_eq_total_bosonic_Hamiltonian} becomes
\begin{align}
    U(t) = e^{-iH t} = \ket{0_S}\bra{0_S} \otimes e^{-i\left(\frac{\omega_0}{2} + H_+\right)t} + \ket{1_S}\bra{1_S}\otimes e^{-i\left(-\frac{\omega_0}{2} + H_-\right)t},
\end{align}
where 
\begin{align}
    H_\pm = H_B \pm \sum_k g_k (a_k^\dagger + a_k),
\end{align}
and $\sigma^z\ket{0_S} = \ket{0_S}$, $\sigma^z\ket{1_S} = -\ket{1_S}$. Applying $\rho_S(t) = \Tr_B\left[U(t)\rho_{SB}(0)U(t)^\dagger\right]$, it can be shown that the diagonal elements of the system's density matrix remain unchanged, and the off-diagonal elements evolve as 
\begin{align}
    \rho_{01}(t) = \rho_{01}(0)e^{-i\omega_0 t}\chi(t),
\end{align}
where $\chi(t) = \bra{0_B}e^{iH_-t}e^{-iH_+ t}\ket{0_B}$. For the $k$-th mode of the bosonic bath, we can write a displacement operator of the form $D_k(\alpha) = e^{\alpha a^\dagger_k - \alpha a_k}$, which satisfies $D_k(\alpha)a_kD_k(\alpha)^\dagger = a_k + \alpha$, where $\alpha_k = \frac{g_k}{\omega_k}$. Using this displacement operator $D_k(\alpha)$, $H_{\pm,k}$ are given by 
\begin{align}
    H_{+, k} = D_k(\alpha)\left(\omega_k a_k^\dagger a_k\right) D_k(\alpha)^\dagger - \frac{g_k^2}{\omega_k}, && \text{and} && H_{-, k} = D_k(-\alpha) \left(\omega_k a_k^\dagger a_k\right) D(-\alpha)^\dagger - \frac{g_k^2}{\omega_k}.
\end{align}
Now, the time evolution of the vacuum for mode $k$ is given by 
\begin{align}
    \ket{\psi_{+,k}(t)} &= e^{-iH_{+, k} t}\ket{0_k} = e^{iE_k t}D_k(\alpha_k)e^{-i \left(\omega_k a_k^\dagger a_k\right)t}D_k(\alpha_k)^\dagger \ket{0_k}, \nonumber \\
    \ket{\psi_{-,k}(t)} &= e^{-iH_{-, k} t}\ket{0_k} = e^{iE_k t}D_k(-\alpha_k)e^{-i \left(\omega_k a_k^\dagger a_k\right)t}D_k(-\alpha_k)^\dagger \ket{0_k},
\end{align}
where $E_k = g_k^2/\omega_k$. From above, after some simplification, we can write 
\begin{align}
\braket{\psi_{-, k}(t)|\psi_{+, k}(t)} = \exp\left[-4\frac{g_k^2}{\omega_k^2}\left\{1 - \cos\left(\omega_k t\right)\right\}\right], 
\end{align}
and thus, the factor $\chi(t)$ becomes
\begin{align}
    \chi(t) = \prod_k \exp\left[-4\frac{g_k^2}{\omega_k^2}\left\{1 - \cos\left(\omega_k t\right)\right\}\right] = \exp\left[-4\sum_k \frac{g_k^2}{\omega_k^2}\left\{1 - \cos\left(\omega_k t\right)\right\}\right]. 
\end{align}
Introducing spectral density of the form $J(\omega) = \sum_kg_k^2 \delta(\omega - \omega_k)$, and replacing the sum by an integral in the continuum limit, we get 
\begin{align}\label{SM_eq_chit}
    \chi(t) = \exp\left[-4\int_0^\infty d\omega \frac{J(\omega)}{\omega^2}\left\{1 - \cos(\omega t)\right\}\right] = \exp\left[-8 \int_0^\infty d\omega \frac{J(\omega)}{\omega^2}\sin^2\left(\frac{\omega t}{2}\right)\right].
\end{align}

Having obtained the final form of the dephasing factor $\chi(t)$, we now discuss its incompatibility in producing RTN dynamics in various regimes. There are multiple reasons for this incompatibility. Straightforwardly, we can point one out for the \textit{case A:} $\left(\frac{b}{\gamma}\right)^2>1$ in Eq.~\eqref{SM_eq_RTN_Lambda_final_form}, i.e., $\Lambda(t) = e^{-\gamma t}f(t)$, where $f(t) = \cos(\mu\gamma t) + \frac{1}{\mu}\sin(\mu\gamma t)$. It can be easily verified that the function $f(t)$ has multiple roots $t_n$ as, for real $\mu$, $\tan(\mu\gamma t) = -\mu$ is satisfied multiple times, and hence, $\Lambda(t_n)= 0$ for multiple $t_n$'s. On the other hand, $\chi(t)$ is an exponential of a finite real number, which is greater than zero for all $t$. Hence, the equivalence between the function $\Lambda(t)$ and $\chi(t)$ is not possible in this regime. 

Another reason, applicable to all the regimes, $\left(\frac{b}{\gamma}\right)^2\le 1$ and $\left(\frac{b}{\gamma}\right)^2>1$ in Eq.~\eqref{SM_eq_RTN_Lambda_final_form}, is demonstrated below. The function $\Lambda(t)$ is a unique solution to a second-order linear differential equation with constant coefficients:
\begin{align} \label{SM_eq_ODE_for_Lambda2}
    \frac{\der^2\Lambda(t)}{\der t^2} + 2\gamma\frac{\der \Lambda (t)}{\der t} + b^2\Lambda(t) = 0,
\end{align}
with initial conditions $\Lambda(0) = 1$ and $\frac{\der \Lambda(0)}{\der t} = 0$, see above Eq.~\eqref{SM_eq_RTN_Lambda_final_form}. For the sake of contradiction, let us assume that the factor $\chi(t)$ is equivalent to $\Lambda(t)$ for some choice of $J(\omega)$. If this were true, $\chi(t) = e^{-F(t)}$ would also have to satisfy the same differential equation for some positive constants $b$ and $\gamma$. To investigate this, we calculate the first and second derivatives of the function $\chi(t)$,
\begin{align}
    \frac{\der \chi(t)}{\der t} = -F'(t)e^{-F(t)} = -F'(t)\chi(t), && \text{and} && \frac{\der^2 \chi(t)}{\der t^2} = \left[-F''(t) + \left(F'(t)\right)^2\right]e^{-F(t)} = \left[-F''(t) + \left(F'(t)\right)^2\right]\chi(t),
\end{align}
and make a differential equation similar to Eq.~\eqref{SM_eq_ODE_for_Lambda2} by replacing $\Lambda(t)$ with $\chi(t)$, that is,
\begin{align}
    \left[-F''(t) + \left(F'(t)\right)^2\right]\chi(t) - 2\gamma F'(t)\chi(t) + b^2 \chi(t) = 0. 
\end{align}
Since $\chi(t)$ is not identically zero, we can divide the above equation by $\chi(t)$ and rearrange the terms to get 
\begin{align}\label{SM_eq_Ft_ode}
    F''(t) - (F'(t))^2 + 2\gamma F'(t) = b^2.
\end{align}
On comparing $\chi(t) = e^{-F(t)}$ with Eq.~\eqref{SM_eq_chit}, we get $F(t) = 4\int_0^\infty d\omega \frac{J(\omega)}{\omega^2}(1 - \cos\left(\omega t\right))$, using which we get $F'(t) = 4\int_0^\infty d\omega \frac{J(\omega)}\omega \sin(\omega t)$ and $F''(t) = 4\int_0^\infty d\omega J(\omega) \cos(\omega t)$. These factors can be substituted in the above equation to get
\begin{align}\label{SM_eq_contradiction}
    4\int_0^\infty d\omega J(\omega) \cos(\omega t) - \left(4\int_0^\infty d\omega \frac{J(\omega)}\omega \sin(\omega t)\right)^2 + 8\gamma \int_0^\infty d\omega \frac{J(\omega)}\omega \sin(\omega t) = b^2.
\end{align}
For any non-trivial and practical spectral density $J(\omega)$, the left-hand side of the above equation is inherently a function of time $t$, whereas the right-hand side is a constant $b^2$. This contradicts our assumption that $\chi(t)$ can be a solution of the ordinary differential equation~\eqref{SM_eq_ODE_for_Lambda2}, making $\chi(t)$ essentially different from $\Lambda(t)$ for all the regimes.

As a particular example, the above arguments can be explicitly shown to hold for the Ohmic spectral density. Consider the ohmic spectral density $J(\omega) = A \omega e^{-\omega/\Omega}$, where $A$ is a dimensionless coupling constant and $\Omega$ is an upper cut-off frequency. The factors $F'(t)$ and $F''(t)$ for this spectral density are
\begin{align}
    F'(t) = 4A\int_0^\infty d\omega e^{-\omega/\Omega}\sin(\omega t), && \text{and} && F''(t) = 4A\int_0^\infty d\omega e^{-\omega/\Omega} \omega\cos(\omega t).  
\end{align}
The above integrations can be evaluated to be
\begin{align}
    F'(t) = \frac{4A\Omega^2 t}{1 + \Omega^2 t^2}, && \text{and} && F''(t) = \frac{4A \Omega^2 (1 - \Omega^2t^2)}{\left(1 + \Omega^2 t^2\right)^2},
\end{align}
such that the left-hand side of Eq.~\eqref{SM_eq_Ft_ode} is
\begin{align}
    F''(t) - (F'(t))^2 + 2\gamma F'(t) = \frac{4A \Omega^2\left(1 + 2t\gamma - 4At^2\Omega^2 + 2t^3\gamma\Omega^2 - t^4\Omega^4\right)}{1 + t^2\Omega^2}^2.
\end{align}
The above expression is evidently a rational function of time $t$. For it to be equal to $b^2$, it must always have the same value for all $t$, which is evidently not the case. Consequently, Eqs.~\eqref{SM_eq_Ft_ode} and~\eqref{SM_eq_contradiction} are not applicable for the Ohmic spectral density. A similar exercise can be performed with other spectral densities, yielding a similar result. Hence, a quantum system's interaction with the usual bosonic bath doesn't lead to an RTN evolution. 

\end{document}